\documentclass[a4paper,12pt,twoside]{article}
\usepackage{amsmath}
\usepackage{mathtools}

\usepackage[T1]{fontenc}
\usepackage{mathptmx}
\usepackage{graphicx}
\usepackage{hyperref}

\usepackage{multirow}
\usepackage{natbib}

\usepackage{geometry}
\usepackage{fancyhdr}
\usepackage{titlesec}
\titleformat*{\section}{\large\bfseries}
\titleformat*{\subsection}{\bfseries}

\usepackage{listings}

\makeatletter
  
  \@addtoreset{equation}{section}
\makeatother

\hypersetup{
  setpagesize = false,
  bookmarksnumbered = true,
  bookmarksopen = true,
  hidelinks
}

\begin{document}

\thispagestyle{headings}

\noindent
\vspace{24pt}
\hrule
\begin{center}
  {\Large
    \textbf{Systematic simulation of age-period-cohort analysis: \\
    Demonstrating bias of Bayesian regularization}
  }
\end{center}
\hrule

\vspace{36pt}
\begin{center}
  Yuta Matsumoto\textsuperscript{\dag}\\
  \vspace{12pt}
  \textsuperscript{\dag}
  Quality Assurance Office Institutional Research,
  Hosei University, Japan \\
  \texttt{yuta.matsumoto.76@adm.hosei.ac.jp} \\
  \vspace{12pt}
  October 11, 2023
\end{center}

\vspace{24pt}
\begin{abstract}
Age-period-cohort (APC) analysis is
one of the fundamental time-series analyses
used in the social sciences.
It is well known that APC analysis has an identification problem,
and applying multilevel analysis results in
the linear component of the cohort effects close to zero.
However, previous studies do not compare other overcoming methods,
such as the intrinsic estimator and the Bayesian cohort model,
by simulation of various patterns.
This paper evaluates APC analysis via systematic simulation in term of
how well the artificial parameters are recovered.
We consider three models of Bayesian regularization
using normal prior distributions:
the random effects model with reference to multilevel analysis,
the ridge regression model equivalent to the intrinsic estimator,
and the random walk model referred to as the Bayesian cohort model.
The proposed simulation generates artificial data
through combinations of the linear components,
focusing on the fact that the identification problem affects
the linear components of the three effects.
Among the 13 cases of artificial data,
the random walk model recovered the artificial parameters well in 10 cases,
while the random effects model and the ridge regression model did so in 4 cases.
The cases in which the models failed to recover the artificial parameters show
the estimated linear component of the cohort effects as close to zero.
In conclusion,
the models of Bayesian regularization in APC analysis have a bias:
the index weights have a large influence on the cohort effects
and these constraints drive the linear component of the cohort effects
close to zero.
However, the random walk model mitigates
underestimating the linear component of the cohort effects.
\end{abstract}

\section{Introduction}

Age-period-cohort (APC) analysis is
one of the fundamental time-series analyses used
in the social sciences.
In APC analysis,
age effects reflect the influence of individual differences in age,
period effects reflect the influence of differences in time period,
and cohort effects represent the influence of differences in birth year.
APC analysis is important
because long-term changes are also the result of demographic metabolism,
in which older generations leave society
and younger generations with different characteristics enter
\citep{Ryder1965}.
Given the nature of such cohort replacement,
we need to consider not only period effects
but also cohort effects.

It is well-known that
APC analysis has a serious issue of identification.
The linear components of the three effects are mixed
due to the linear dependence of the factors
when cohort is linearly associated with age and period
according to the relationship cohort $=$ period $-$ age.
In other words,
the APC identification problem makes it impossible
to directly estimate the linear components of these effects.
Although many previous studies have applied multilevel analysis
to solve the rank deficiency of the design matrix,
this constraint results in the linear component of the cohort effect
being close to zero \citep{Fosse2019}.
In addition, \citet{Sakaguchi2019} mathematically derived
that the Bayesian cohort model has less bias for cohort effects
than multilevel analysis.
However, these constraints are still controversial \citep{O'Brien2014},
as the above previous studies do not compare
the Bayesian cohort model with the intrinsic estimator,
another well-known method in APC analysis \citep{Yang2013},
and do not evaluate the performance of the three models
via simulation of various patterns.

The present paper proposes a simulation
that systematically verifies APC analysis.
The proposed simulation generates artificial data
through combinations of the linear components,
focusing on the fact that the identification problem affects
the linear components of the three effects.
We consider three models of Bayesian regularization
using normal prior distributions:
the random effects model with reference to multilevel analysis,
the ridge regression model equivalent to the intrinsic estimator,
and the random walk model referred to as the Bayesian cohort model.
We execute the three models
using the simulation-generated artificial data
and evaluate the models in terms of how well
the artificial parameters are recovered.

This paper reviews APC analysis from Sections 2 to 4.
Specifically, Section 2 shows the notation of APC analysis
and the identification problem.
Section 3 describes
the constraints for the random effects model,
the ridge regression model, and the random walk model
through the framework of Bayesian regularization.
Section 4 refers to the linear component,
which is the basis of the identification problem,
and shares the mathematical evidence
that the random walk model can be expected
to perform better than the random effects model,
as clarified by the previous study.
Section 5 presents the systematic simulation proposed in this paper
and the definition of the bias evaluation function,
and verifies the performance of the three models
by systematic simulations.
Section 6 concludes that there is no one-size-fits-all APC analysis,
but that the random walk model performs relatively well.

\section{Age-period-cohort analysis}

\subsection{Notation}

Let $i=1, \ldots I$ denote the index of the age group,
$j=1, \ldots J$ denote the index of the period group,
$k=1, \ldots K$ denote the index of the cohort group,
and these three indexes be determined by
\begin{equation}
  k=j-i+I, \quad K=I+J-1,
    \label{Eq2:Relationship_Index_IJK}
\end{equation}
if the intervals of age and period have the same scale.
The general model for APC analysis is
\begin{equation}
  y_{i,j} = b_{0} + b_{i}^{A} + b_{j}^{P} + b_{k}^{C} +
    \epsilon_{i,j},
    \label{Eq2:BasicModel_APC}
\end{equation}
where $y_{i,j}$ denotes the observed value
(see Table \ref{Tb2:ObservedValue}),
$b_{0}$ denotes the intercept,
$b_{i}^{A}$ denotes the age effect,
$b_{j}^{P}$ denotes the period effect,
$b_{k}^{C}$ denotes the cohort effect,
$\epsilon_{i,j}$ denotes the error term,
and each effect satisfies the sum-to-zero condition,
\begin{equation*}
  \sum_{i=1}^{I} b_{i}^{A} = \sum_{j=1}^{J} b_{j}^{P} =
    \sum_{k=1}^{K} b_{k}^{C} = 0.
\end{equation*}

\begin{table}[t]
  \centering
  \caption{Observed values in an age-period table}
  \label{Tb2:ObservedValue}
  \begin{normalsize}
  \begin{tabular}{c|ccc}
    \hline
    $y_{i,j}, \; k$ & $j=1$ & $\cdots$ & $j=J$ \\
    \hline
    $i=1$ & $y_{1,1}, \; k=I$  & $\cdots$ & $y_{1,J}, \; k=K$ \\
    $\vdots$ & $\vdots$ & $\ddots$ & $\vdots$ \\
    $i=I$ & $y_{I,1}, \; k=1$ & $\cdots$ & $y_{I,J}, \; k=J$ \\
    \hline
  \end{tabular}
\end{normalsize}
\end{table}

Here, when the error terms are approximated by
normal distributions,
the model is written as
\begin{align*}
  y_{n} & \sim \mathrm{Normal}
    \left(
      \mu_{n}, \sigma
    \right) \qquad \qquad n=1, \ldots, N, \\
  \mu_{n} & = b_{0} + \sum_{i=1}^{I} x_{n,i}^{A} b_{i}^{A} +
    \sum_{j=1}^{J} x_{n,j}^{P} b_{j}^{P} +
    \sum_{k=1}^{K} x_{n,k}^{C} b_{k}^{C},
\end{align*}
where
$y_{n}$ is $y_{i,j}$ rearranged as the component
of a vector with $N$ rows,
$\sigma$ denotes a standard deviation, and
$x_{n,i}^{A}$, $x_{n,j}^{P}$, and $x_{n,k}^{C}$
are the components of the design matrix composed of three factors.
Then, the log likelihood is
\begin{equation}
  \log L = - N \log \sigma - \frac{1}{2 \sigma^{2}} \sum_{n=1}^{N}
    \left(
      y_{n} - \mu_{n}
    \right)^{2},
    \label{Eq2:LogLikelihood_LinearModel}
\end{equation}
excluding the constant term.
In general, estimates are obtained by maximizing $\log L$
(\ref{Eq2:LogLikelihood_LinearModel});
however, we need to add constraints in APC analysis
since it is not possible to uniquely determine the estimates
owing to the identification problem described below.

\subsection{Identification problem}

To understand the identification problem,
it is convenient to center each index\footnote{
  $v_{i}^{A}$, $v_{j}^{P}$, and $v_{k}^{C}$ are
  the components of the null vector of
  the design matrix composed of the three factors.
} \citep{Kupper1985},
\begin{equation*}
  v_{i}^{A} = i - \frac{I+1}{2}, \quad
    v_{j}^{P} = j - \frac{J+1}{2}, \quad
    v_{k}^{C} = k - \frac{K+1}{2}.
\end{equation*}
Here, the equation
\begin{equation*}
  v_{i}^{A} - v_{j}^{P} + v_{k}^{C} = 0,
\end{equation*}
is satisfied
using the relationship of the cohort index
(\ref{Eq2:Relationship_Index_IJK}).
Thus, the right-hand side of the model (\ref{Eq2:BasicModel_APC}) becomes
\begin{equation*}
  b_{0} + b_{i}^{A} + b_{j}^{P} + b_{k}^{C} + \epsilon_{i,j}
  =
  b_{0} + b_{i}^{A} + b_{j}^{P} + b_{k}^{C} +
    \left(
      v_{i}^{A} - v_{j}^{P} + v_{k}^{C}
    \right) + \epsilon_{i,j},
\end{equation*}
and we can write the general solutions of the three effects as
\begin{equation}
  b_{i}^{A} = \widehat{b}_{i}^{A} + s v_{i}^{A}, \quad
    b_{j}^{P} = \widehat{b}_{j}^{P} - s v_{j}^{P}, \quad
    b_{k}^{C} = \widehat{b}_{k}^{C} + s v_{k}^{C},
    \label{Eq2:GeneralSolution_APC}
\end{equation}
where $\widehat{b}_{i}^{A}$, $\widehat{b}_{j}^{P}$,
and $\widehat{b}_{k}^{C}$ are
the particular solutions of the three effects
and $s$ denotes an arbitrary real number.

In summary, the APC identification problem is that
there are many maximum likelihood estimates of the model
(\ref{Eq2:LogLikelihood_LinearModel})
owing to the linear dependency of cohort $=$ period $-$ age.
Since $v_{i}^{A}$, $v_{j}^{P}$, and $v_{k}^{C}$
constitute the linear components of the three effects,
they are offset completely
when the slopes of the age and cohort effects increase by $s$
and the slope of the period effect decreases by $s$.
As a result, the linear components of the three effects
are mixed in the observed values,
while the nonlinear components
that are not affected by the identification problem
can be easily separated.

\section{Bayesian regularization}

To overcome the identification problem,
Bayesian regularization constrains the parameters of the three effects
by assuming prior probabilities.
It is a strategy to statistically estimate
mathematically indistinguishable linear components
by using mathematically identifiable nonlinear components and priors.
If there are an infinite number of maximum likelihood estimates,
as in APC analysis,
point estimates that maximize the posterior probabilities of Bayesian models
are determined by maximizing the priors.

\subsection{Random effects model}

Many studies use a multilevel analysis
that reflects the nesting of individuals
in groups of period and cohort \citep{Yang2006},
but treat the age effects as fixed effects,
which means that the age effects are unconstrained.
In this paper, we consider the random effects model
with reference to multilevel analysis,
where the model assumes a normal distribution for
the prior probabilities of each of the three effects.
The priors are
\begin{align*}
  b_{i}^{A}  & \sim \mathrm{Normal} \,
    (0, \sigma^{A} ) & i &  = 1,
    \ldots , I, \\
  b_{j}^{P} & \sim \mathrm{Normal} \,
    (0, \, \sigma^{P} ) & j & = 1,
    \ldots , J,  \\
  b_{k}^{C} & \sim \mathrm{Normal} \,
    (0, \sigma^{C} ) & k &  = 1,
    \ldots , K,
\end{align*}
where $\sigma^{A}$, $\sigma^{P}$, and $\sigma^{C}$
denote the standard deviations of the three effects
and the log priors are
\begin{equation}
  \begin{split}
    \log RE
    & = -
    \left (
      I \log \sigma^{A} + J \log \sigma^{P} +
      K \log \sigma^{C}
    \right )
    \\
    & \qquad - \frac{1}{2}
    \left \{
      \frac{1}{ (\sigma^{A})^{2} } \sum_{i=1}^{I} (b_{i}^{A})^{2} +
      \frac{1}{ (\sigma^{P})^{2} } \sum_{j=1}^{J} (b_{j}^{P})^{2} +
      \frac{1}{ (\sigma^{C})^{2} } \sum_{k=1}^{K} (b_{k}^{C})^{2}
    \right \},
  \end{split}
  \label{Eq3:LogPrior_RE}
\end{equation}
excluding the constant term.
Here, maximizing $\log RE$ (\ref{Eq3:LogPrior_RE})
means minimizing the sum of squares of the parameters.

\subsection{Ridge regression model}

Ridge regression analysis is
a method that imposes the sum of squares of the parameters
as a penalty
and is designed to overcome the adverse effects of multicollinearity.
The ridge regression model in this paper is implemented
by assuming normal distributions with zero means
and equal standard deviations
for the prior probabilities of the three effects.
Unifying the standard deviations,
\begin{equation}
  \lambda = \sigma^{A} = \sigma^{P} = \sigma^{C},
    \label{Eq3:StandardDeviation_RR}
\end{equation}
and substituting (\ref{Eq3:StandardDeviation_RR}) into
$\log RE$ (\ref{Eq3:LogPrior_RE}),
we can write the log priors as
\begin{equation}
  \log RR = - (I+J+K) \log \lambda
    - \frac{1}{2 \lambda^2}
    \left \{
      \sum_{i=1}^{I} (b_{i}^{A})^{2} +
      \sum_{j=1}^{J} (b_{j}^{P})^{2} +
      \sum_{k=1}^{K} (b_{k}^{C})^{2}
    \right \}.
    \label{Eq3:LogPrior_RR}
\end{equation}
Here, maximizing $\log RR$ (\ref{Eq3:LogPrior_RR})
means minimizing the sum of squares of the parameters
as well as $\log RE$ (\ref{Eq3:LogPrior_RE}).

The intrinsic estimator is another well-known method in APC analysis
and produces similar results to the ridge regression model,
as this operation minimizes the Euclidean norm of the parameters,
giving a particular solution
that is the average of the general solution \citep{Yang2004}.

\subsection{Random walk model}

We can also apply time-series models to APC analysis
based on a previous study
that proposes smoothing the cohort effects \citep{Fu2008}.
The random walk model literally assumes
a random walk for the prior probabilities of the three effects
\citep{Schmid2007} and can be written as
\begin{align*}
  b_{i+1}^{A} & \sim \mathrm{Normal} \,
    (b_{i}^{A}, \sigma^{A} ) & i & = 1,
    \ldots , I-1, \\
  b_{j+1}^{P} & \sim \mathrm{Normal} \,
    (b_{j}^{P}, \sigma^{P} ) & j & = 1,
    \ldots , J-1,  \\
  b_{k+1}^{C} & \sim \mathrm{Normal} \,
    (b_{k}^{C}, \sigma^{C} ) & k & = 1,
    \ldots , K-1.
\end{align*}
The log priors can be summarized as follows:
\begin{equation}
  \begin{split}
    & \log RW = -
    \left \{
      (I-1) \log \sigma^{A} + (J-1) \log \sigma^{P} +
      (K-1) \log \sigma^{C}
    \right \}
    \\
    & \quad - \frac{1}{2}
    \left\{
      \frac{1}{ (\sigma^{A})^{2} } \sum_{i=1}^{I-1} (b_{i+1}^{A} - b_{i}^{A})^{2}
     + \frac{1}{ (\sigma^{P})^{2} } \sum_{j=1}^{J-1} (b_{j+1}^{P} - b_{j}^{P})^{2} +
      \frac{1}{ (\sigma^{C})^{2} } \sum_{k=1}^{K-1} (b_{k+1}^{C} - b_{k}^{C})^{2}
    \right\},
  \end{split}
  \label{Eq3:LogPrior_RW}
\end{equation}
excluding the constant term.

Here, maximizing $\log RW$ (\ref{Eq3:LogPrior_RW})
means, unlike $\log RE$ and $\log RR$,
minimizing the sum of squares of
the differences in the adjacent parameters.
Furthermore, the random walk model is equivalent to
the Bayesian cohort model proposed by \citet{Nakamura1986}
and this constraint takes advantage of the fact
that age, period, and cohort indexes are ordered.

\section{Mathematical mechanism of bias}

\subsection{Linear and nonlinear components}

APC analysis depends heavily on the way in which
the constraints assign the linear components to the three effects.
Thus, we separate the linear and nonlinear components
of the general solution
as $b_{i}^{A}= b_{i}^{A[\mathit{L}]}+b_{i}^{A[\mathit{NL}]}$
\citep{Sakaguchi2019}
in order to discuss constraint bias.
For example,
we regress the particular solution of the age effects on the centering index
and let $s^{A}$ denote the obtained slope.
The equation is
$\widehat{b}_{i}^{A} = s^{A} v_{i}^{A} + \epsilon_{i}^{A}$ and
$\epsilon_{i}^{A}=b_{i}^{A[\mathit{NL}]}$,
as $\epsilon_{i}^{A}$ does not contain the linear component.
Here, the particular solutions of the three effects are
\begin{equation*}
  \widehat{b}_{i}^{A} = s^{A} v_{i}^{A} + b_{i}^{A[\mathit{NL}]}, \quad
    \widehat{b}_{j}^{P} = s^{P} v_{j}^{P} + b_{j}^{P[\mathit{NL}]}, \quad
    \widehat{b}_{k}^{C} = s^{C} v_{k}^{C} + b_{k}^{C[\mathit{NL}]},
\end{equation*}
where $s^{P}$ and $s^{C}$ are the slopes calculated
from the particular solutions of the period and cohort effects.
By substituting the above solutions into
(\ref{Eq2:GeneralSolution_APC}),
the general solutions can be rewritten as follows:
\begin{equation}
  b_{i}^{A} =  (s^{A} + s) \,  v_{i}^{A} + b_{i}^{A[\mathit{NL}]}, \quad
    b_{j}^{P} = (s^{P} - s) \, v_{j}^{P} + b_{j}^{P[\mathit{NL}]}, \quad
    b_{k}^{C} = (s^{C} + s) \, v_{k}^{C} + b_{k}^{C[\mathit{NL}]}.
    \label{Eq4:GeneralSolution_APC_Separate_NonLinear}
\end{equation}
Therefore,
the linear components of the general solutions are expressed as
\begin{equation}
  b_{i}^{A[\mathit{L}]} = (s^{A} + s) \, v_{i}^{A}, \quad
  b_{j}^{P[\mathit{L}]} = (s^{P} - s) \, v_{j}^{P}, \quad
  b_{k}^{C[\mathit{L}]} = (s^{C} + s) \, v_{k}^{C},
  \label{Eq4:GeneralSolution_LinearComponent}
\end{equation}
using the centering indexes.

\subsection{Linear components represented by indexes}

We can rewrite the linear components of the general solutions
for the models of Bayesian regularization using the centering indexes.
Here, the equation
\begin{equation*}
  \sum_{i=1}^{I}
    b_{i}^{A\mathit{[L]}} b_{i}^{A[\mathit{NL}]}
    =
    (s^{A} + s)
    \sum_{i=1}^{I} v_{i}^{A} b_{i}^{A[\mathit{NL}]} = 0,
\end{equation*}
is satisfied
since the linear and nonlinear components are orthogonal,
and the sum of squares of the parameters is
\begin{gather*}
  \sum_{i=1}^{I} (b_{i}^{A})^{2}
    = \sum_{i=1}^{I}
      \left \{
        (s^{A} + s) \, v_{i}^{A} + b_{i}^{A[\mathit{NL}]}
      \right \}^{2}
    = (s^{A} + s)^{2} \sum_{i=1}^{I} (v_{i}^{A})^{2}
    + \sum_{i=1}^{I} (b_{i}^{A[\mathit{NL}]})^{2},
\end{gather*}
using the general solutions
(\ref{Eq4:GeneralSolution_APC_Separate_NonLinear}).
Therefore,
the log priors of the random effects model
(\ref{Eq3:LogPrior_RE})
can be rewritten as
\begin{gather}
  \log RE^{A} = -I \log \sigma^{A} - \frac{1}{2 (\sigma^{A})^{2}}
    \left \{
      (s^{A} + s)^{2} \sum_{i=1}^{I} (v_{i}^{A})^{2}
      + \sum_{i=1}^{I} (b_{i}^{A[\mathit{NL}]})^{2}
    \right \}, \notag \\
  \log RE^{P} = -J \log \sigma^{P} - \frac{1}{2 (\sigma^{P})^{2}}
    \left \{
      (s^{P} - s)^{2} \sum_{j=1}^{J} (v_{j}^{P})^{2}
      + \sum_{j=1}^{J} (b_{j}^{P[\mathit{NL}]})^{2}
    \right \}, \notag \\
  \log RE^{C} = -K \log \sigma^{C} - \frac{1}{2 (\sigma^{C})^{2}}
    \left \{
      (s^{C} + s)^{2} \sum_{k=1}^{K} (v_{k}^{C})^{2}
      + \sum_{k=1}^{K} (b_{k}^{C[\mathit{NL}]})^{2}
    \right \}, \notag \\
  \log RE = \log RE^{A} + \log RE^{P} + \log RE^{C}.
  \label{Eq4:LogPrior_RE_byIndex}
\end{gather}
The log priors of the ridge regression model are obtained by substituting
the unified standard deviation (\ref{Eq3:StandardDeviation_RR}) into
the log prior of the random effects model
(\ref{Eq4:LogPrior_RE_byIndex}).

Next, the sum of squares of
the differences in the adjacent parameters is
\begin{align*}
  & \sum_{i=1}^{I-1} (b_{i+1}^{A} - b_{i}^{A})^{2}
    = \sum_{i=1}^{I-1}
    \left \{
      (s^{A} + s) + (b_{i+1}^{A[\mathit{NL}]} - b_{i}^{A[\mathit{NL}]})
    \right \}^{2} \\
  & \quad = (s^{A} + s)^{2} (I-1) + 2 (s^{A} + s)
    (b_{I}^{A[\mathit{NL}]} - b_{1}^{A[\mathit{NL}]}) +
    \sum_{i=1}^{I-1} (b_{i+1}^{A[\mathit{NL}]} - b_{i}^{A[\mathit{NL}]})^{2},
\end{align*}
using the general solutions
(\ref{Eq4:GeneralSolution_APC_Separate_NonLinear}).
Thus, the log priors of the random walk model
(\ref{Eq3:LogPrior_RW})
can be rewritten as follows
\begin{gather}
  \begin{split}
    \log RW^{A} =
    &  - (I-1) \log \sigma^{A} - \frac{1}{2(\sigma^{A})^{2}}
      \Biggl\{ (s^{A} + s)^{2} (I-1) \\
    & \qquad \qquad  + 2 (s^{A} + s) (b_{I}^{A[\mathit{NL}]} - b_{1}^{A[\mathit{NL}]}) +
    \sum_{i=1}^{I-1} (b_{i+1}^{A[\mathit{NL}]} - b_{i}^{A[\mathit{NL}]})^{2}
    \Biggr\},
  \end{split} \notag \\
  \begin{split}
    \log RW^{P} =
    & - (J-1) \log \sigma^{P} - \frac{1}{2(\sigma^{P})^{2}}
      \Biggl\{ (s^{P} - s)^{2} (J-1) \\
    & \qquad \qquad + 2 (s^{P} - s) (b_{J}^{P[\mathit{NL}]} - b_{1}^{P[\mathit{NL}]}) +
    \sum_{j=1}^{J-1} (b_{j+1}^{P[\mathit{NL}]} - b_{j}^{P[\mathit{NL}]})^{2}
    \Biggr\},
  \end{split} \notag \\
  \begin{split}
    \log RW^{C} =
    & - (K-1) \log \sigma^{C} - \frac{1}{2(\sigma^{C})^{2}}
      \Biggl\{ (s^{C} + s)^{2} (K-1)  \\
    & \qquad \qquad + 2 (s^{C} + s) (b_{K}^{C[\mathit{NL}]} - b_{1}^{C[\mathit{NL}]}) +
    \sum_{k=1}^{K-1} (b_{k+1}^{C[\mathit{NL}]} - b_{k}^{C[\mathit{NL}]})^{2}
    \Biggr\},
  \end{split} \notag \\
  \log RW = \log RW^{A} + \log RW^{P} + \log RW^{C}.
  \label{Eq4:LogPrior_RW_byIndex}
\end{gather}

\subsection{Index weights of linear components}

The linear components cause the difference in the estimates
and are weighted by the indexes in the general solutions
(\ref{Eq4:GeneralSolution_LinearComponent})
while the nonlinear components are uniquely determined
by the observed values.
We roughly describe the bias of the linear components
by fixing $\sigma^{A} = \sigma^{P} = \sigma^{C} = 1$
so that the influence of the nonlinear components is not considered.
The log priors of the random effects model
(\ref{Eq4:LogPrior_RE_byIndex})
can be represented as
\begin{equation*}
  \begin{split}
    \log RE =
    & - \frac{1}{2} \Biggl\{
      (s^{A} + s)^{2} \sum_{i=1}^{I} (v_{i}^{A})^{2} +
      (s^{P} - s)^{2} \sum_{j=1}^{J} (v_{j}^{P})^{2} \\
    & \qquad \qquad + (s^{C} + s)^{2} \sum_{k=1}^{K} (v_{k}^{C})^{2}
      + (\text{the nonlinear components}) \Biggr\},
  \end{split}
\end{equation*}
excluding the constant term.
Here, maximizing $\log RE$
is achieved by $s$
where $(s^{C} + s)^{2}$ is smaller than
$(s^{A} + s)^{2}$ and $(s^{P} - s)^{2}$
because $K=I+J-1$ (\ref{Eq2:Relationship_Index_IJK})
makes the index weights
$\sum_{k=1}^{K} (v_{k}^{C})^{2} > \sum_{i=1}^{I} (v_{i}^{A})^{2}$ and
$\sum_{k=1}^{K} (v_{k}^{C})^{2} > \sum_{j=1}^{J} (v_{j}^{P})^{2}$.
In other words, the index weights exert a strong pressure
to shrink the linear component of the cohort effects;
consequently, they tend to be flat.
The above also occurs with the ridge regression model.
The log priors of the random walk model
 (\ref{Eq4:LogPrior_RW_byIndex})
can be represented as
\begin{equation*}
  \begin{split}
    \log RW =
    & - \frac{1}{2} \Biggl\{
      (s^{A} + s)^{2} (I-1) +
      (s^{P} - s)^{2} (J-1) \\
    & \qquad \qquad + (s^{C} + s)^{2} (K-1)
      + (\text{the nonlinear components}) \Biggr\},
  \end{split}
\end{equation*}
excluding the constant term.
For the same reason,
it also tends to underestimate
the linear component of the cohort effects owing to the index.

Bayesian regularizations
using normal distributions have in common
that the linear component of the cohort effects is close to zero.
However, \citet{Sakaguchi2019} suggested
that the random walk model performs well compared
to the random effects model;
consequently, this paper examines the impact of
the index weights.
Focusing on period and cohort,
the ratio is $J-1:K-1$ for the random walk model and
$\sum_{j=1}^{J} (v_{j}^{P})^{2}: \sum_{k=1}^{K} (v_{k}^{C})^{2}$
for the other two models.
The squared sums of the centering index are
\begin{equation*}
  \sum_{j=1}^{J} (v_{j}^{P})^{2} = \frac{1}{12} \, J \, (J+1) (J-1), \quad
  \sum_{k=1}^{K} (v_{k}^{C})^{2} = \frac{1}{12} \, K \, (K+1) (K-1).
\end{equation*}
Here, a comparison of the above ratios of the index weights shows
\begin{equation}
  \frac{\sum_{k=1}^{K} (v_{k}^{C})^2}{\sum_{j=1}^{J} (v_{j}^{P})^2}
    - \frac{K-1}{J-1}
    = \frac{(K-1) (2J+I) (I-1) }{J \, (J+1) (J-1)} > 0.
    \label{Eq4:Comparison_IndexWeight}
\end{equation}
Thus, the comparison of the index weights
(\ref{Eq4:Comparison_IndexWeight}) suggests
that the random walk model is less affected by the index weights
than are the random effects and ridge regression models.
In other words,
minimizing the sum of squares of the differences in the adjacent parameters
rather than the parameters themselves
mitigates the underestimation of the linear component of the cohort effects.
In the next section, we confirm the performance of the three models
by discussing not only linear components but also nonlinear components.

\section{Simulation}

\subsection{Artificial parameters and data}

\begin{figure}[tp]
  \centering
  \includegraphics[height=22.5truecm]
    {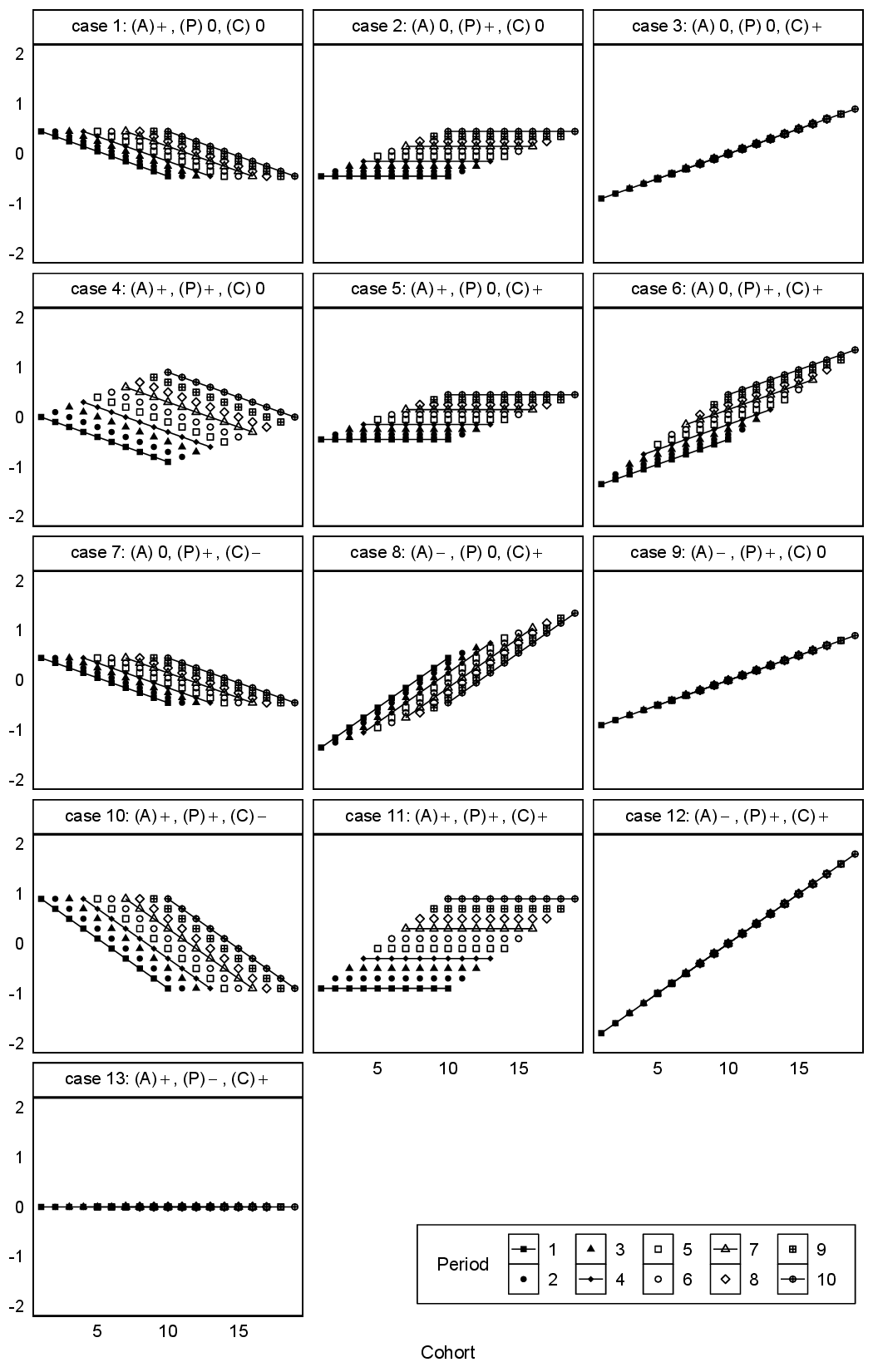}
  \caption{13 cases of artificial data (linear components only)}
  \label{Fg5:ArtificialData_OnlyLinear}
\end{figure}

\begin{figure}[tp]
  \centering
  \includegraphics[height=22.5truecm]
    {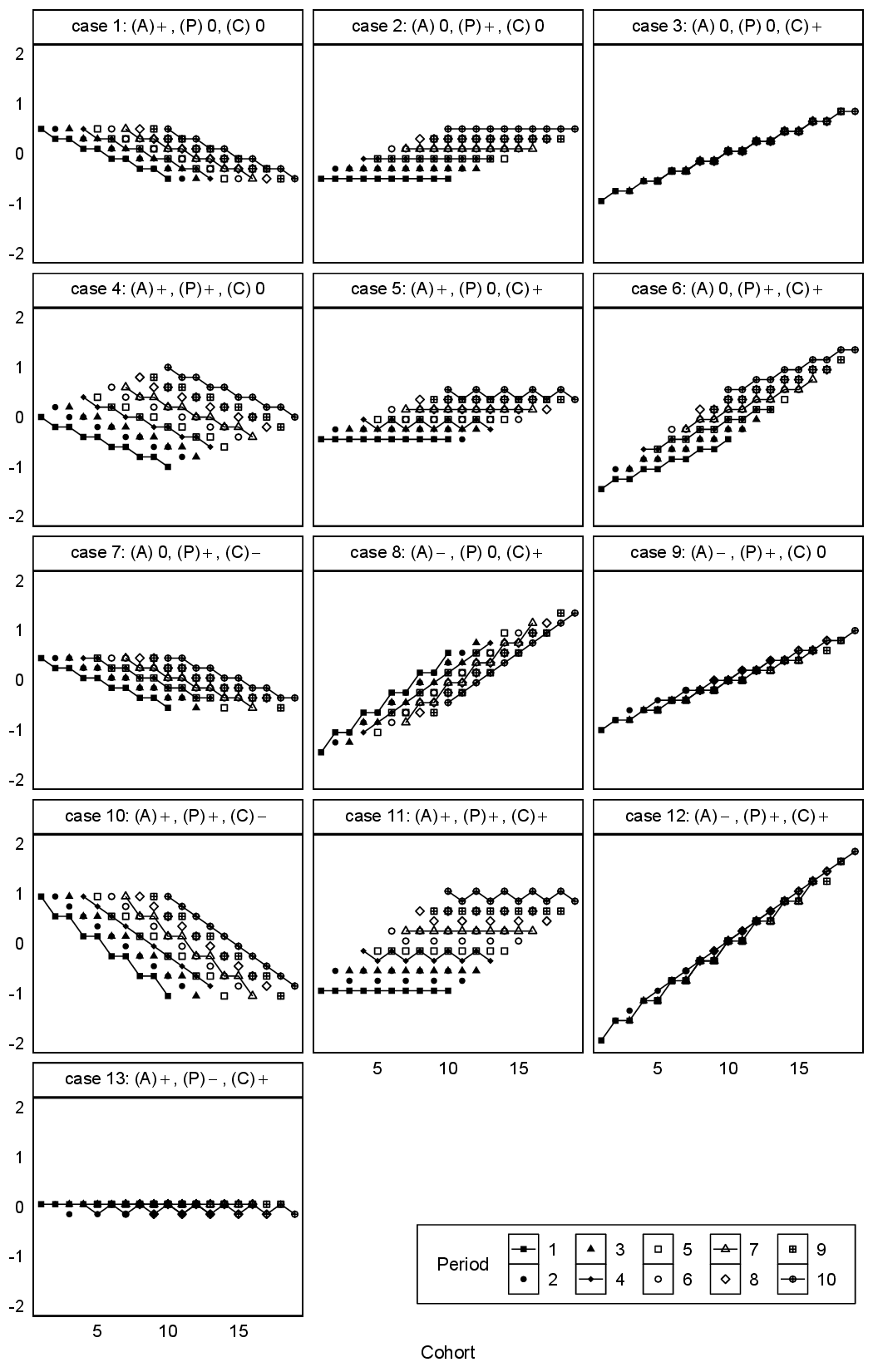}
  \caption{13 cases of artificial data (including nonlinear components)}
  \label{Fg5:ArtificialData_IncludingNonlinear}
\end{figure}

We developed a systematic simulation
to examine the bias of Bayesian regularization.
The objective is to estimate the linear components of the three effects
using nonlinear components and log prior probabilities
in order to overcome the identification problem.
According to our identification strategy,
we add nonlinear components
to the artificial parameters of the factor
containing linear components.
We use trigonometric functions for the nonlinear components,
as they contain no linear components
and can be set to any amount of change.
Accordingly, the artificial parameters
of the three effects can be written as follows:
\begin{align*}
  \beta_{i}^{A} & = - \frac{ \beta^{A[\mathit{NL}]} }{2I}
    \{ \cos (\pi I) -1 \} + \beta^{A[\mathit{L}]} v_{i}^{A} +
    \beta^{A[\mathit{NL}]} \cos (\pi i), \\
  \beta_{j}^{P} & = - \frac{ \beta^{P[\mathit{NL}]} }{2J}
    \{ \cos (\pi J) -1 \} + \beta^{P[\mathit{L}]} v_{j}^{P} +
    \beta^{P[\mathit{NL}]} \cos (\pi j), \\
  \beta_{k}^{C} & = - \frac{ \beta^{C[\mathit{NL}]} }{2K}
    \{ \cos (\pi K) -1 \} + \beta^{C[\mathit{L}]} v_{k}^{C} +
    \beta^{C[\mathit{NL}]} \cos (\pi k),
\end{align*}
where $\beta_{i}^{A[\mathit{L}]}$, $\beta_{j}^{P[\mathit{L}]}$,
and $\beta_{k}^{C[\mathit{L}]}$ denote
the slopes of the artificial parameters,
$\beta_{i}^{A[\mathit{NL}]}$, $\beta_{j}^{P[\mathit{NL}]}$, and
$\beta_{k}^{C[\mathit{NL}]}$ denote the amounts of change
in the nonlinear components,
and the sum-to-zero condition is satisfied\footnote{
  We add
  $- \frac{ \beta^{A[\mathit{NL}]} }{2I} \{ \cos (\pi I) -1 \}$
  to each artificial parameter of the age effects to satisfy
  $\sum_{i=1}^{I} \beta_{i}^{A} =0$
  since $\sum_{i=1}^{I} \beta^{A[\mathit{NL}]} \cos (\pi i)
  = - \beta^{A[\mathit{NL}]}$ when $I$ is an odd number.
},
\begin{equation*}
  \sum_{i=1}^{I} \beta_{i}^{A} = \sum_{j=1}^{J} \beta_{j}^{P} =
  \sum_{k=1}^{K} \beta_{k}^{C} = 0.
\end{equation*}
Here, $y_{n}$ represents the artificial data
generated as follows:
\begin{equation*}
  y_{n} \sim \mathrm{Normal}
    \left(
      \sum_{i=1}^{I} x_{n,i}^{A} \beta_{i}^{A} +
      \sum_{j=1}^{J} x_{n,j}^{P} \beta_{j}^{P} +
      \sum_{k=1}^{K} x_{n,k}^{C} \beta_{k}^{C}, \gamma
    \right)
      \qquad n=1, \ldots, N.
\end{equation*}
We then set
$I=10$, $J=10$, $\gamma = 0.1$ and $N=I \times J \times 10 = 1000$,
so that the error terms generated by the normal distributions
do not greatly affect the simulation.

To conduct a systematic simulation,
we discuss a combination of the linear components
that are the basis of the identification problem.
First, we assume three types of slopes for
the artificial parameters: $0$, $+$, and $-$.
The total number of combinations here is $3^3=27$,
since each effect has three patterns.
Specifically,
$\beta^{A[\mathit{L}]} = 0$ is expressed as $(\mathrm{A}) \, 0$,
$\beta^{P[\mathit{L}]} > 0$ as $(\mathrm{P}) \, +$,
and $\beta^{C[\mathit{L}]} < 0$ as $(\mathrm{C}) \, -$.
In fact, we need only consider $(27-1)/2 = 13$ cases
since we are excluding the cases where there is no linear component,
such as $(\mathrm{A}) \, 0, (\mathrm{P}) \, 0, (\mathrm{C}) \, 0$
and where the positive slope is merely reversed to a negative.
The  combination is thus
cases 1 to 3 having a positive linear component in one factor,
cases 4 to 6 having positive linear components in two factors,
cases 7 to 9 having positive and negative linear components in two factors,
and cases 10 to 13 having linear components in all factors.

We set the variation of the slope to $0.1$
and the nonlinear component to $0.05$.
Specifically, $(\mathrm{A}) \, 0$ represents
$\beta^{A[\mathit{L}]} = 0$ and $\beta^{A[\mathit{NL}]} = 0$,
$(\mathrm{P}) \, +$ represents
$\beta^{P[\mathit{L}]} = 0.1$ and $\beta^{P[\mathit{NL}]} = 0.05$,
and $(\mathrm{C}) \, -$ represents
$\beta^{C[\mathit{L}]} = -0.1$ and $\beta^{C[\mathit{NL}]} = -0.05$.
For better understanding,
Figure \ref{Fg5:ArtificialData_OnlyLinear}
includes only the linear components for the 13 cases
and excludes the error term;
Figure \ref{Fg5:ArtificialData_IncludingNonlinear}
includes the nonlinear components.
The dot plots are visualizations by period and
the x-axis represents cohort.
Figure \ref{Fg5:ArtificialData_OnlyLinear}
shows that cases 1 and 7 are identical
and very similar to case 10.
Moreover, cases 2 and 5 are identical and very similar to case 11, while
cases 3 and 9 are identical
and very similar to case 12.
In addition, the linear components of case 13 are offset
and no variation appears in the artificial data.
In other words, the mixture of linear components
in the identification problem means
that combining different linear components can generate
precisely the same data.
Unlike Figure \ref{Fg5:ArtificialData_OnlyLinear},
Figure \ref{Fg5:ArtificialData_IncludingNonlinear}
does not reveal identical data.
Consequently, we verify
whether the models of Bayesian regularization
recover the artificial parameters using this small difference.

\subsection{Results and bias}

\begin{table}[t]
  \centering
  \caption{All results of the three models applying Bayesian regularization}
  \label{Tb5:AllResults_BayesianRegularization}
  \begin{normalsize}
    \begin{tabular}{cccccrccrccrc}
      \hline
      \multirow{2}{*}{case} & \multicolumn{3}{c}{\shortstack{Artifical\\parameters}} &  &
        \multicolumn{2}{c}{\shortstack{Random effects\\model}}  &  &
        \multicolumn{2}{c}{\shortstack{Ridge regression\\model}} &  &
        \multicolumn{2}{c}{\shortstack{Random walk\\model}} \\
      \cline{2-4} \cline{6-7} \cline{9-10} \cline{12-13}
       & (A) & (P) & (C) & & \multicolumn{1}{c}{$s$} & & & \multicolumn{1}{c}{$s$}  & & & \multicolumn{1}{c}{$s$} & \\
      \hline
      1 & $+$ & $0$ & $0$ & & $0.000$ & A & & $-0.012$ & A & & $0.000$ & A \\
      2 & $0$ & $+$ & $0$ & & $0.000$ & A & & $0.012$ & A & & $0.000$ & A \\
      3 & $0$ & $0$ & $+$ & & $-0.099$ & E & & $-0.077$ & D & & $0.000$ & A \\
      \cline{2-13}
      4 & $+$ & $+$ & $0$ & & $0.000$ & A & & $0.000$ & A & & $0.000$ & A \\
      5 & $+$ & $0$ & $+$ & & $-0.100$ & E & & $-0.089$ & E & & $0.000$ & A \\
      6 & $0$ & $+$ & $+$ & & $-0.099$ & E & & $-0.066$ & D & & $-0.001$ & A \\
      \cline{2-13}
      7 & $0$ & $+$ & $-$ & & $0.100$ & E & & $0.089$ & E & & $0.001$ & A \\
      8 & $-$ & $0$ & $+$ & & $-0.099$ & E & & $-0.066$ & D & & $0.001$ & A \\
      9 & $-$ & $+$ & $0$ & & $0.001$ & A & & $0.023$ & B & & $0.000$ & A \\
      \cline{2-13}
      10 & $+$ & $+$ & $-$ & & $0.100$ & E & & $0.078$ & D & & $0.088$ & E \\
      11 & $+$ & $+$ & $+$ & & $-0.100$ & E & & $-0.078$ & D & & $-0.092$ & E \\
      12 & $-$ & $+$ & $+$ & & $-0.100$ & E & & $-0.054$ & C & & $-0.020$ & B \\
      \cline{2-13}
      13 & $+$ & $-$ & $+$ & & $-0.101$ & E & & $-0.101$ & E & & $-0.105$ & E \\
      \hline
      \end{tabular}
  \end{normalsize}
\end{table}

Before confirming the results,
we need to define the bias evaluation function.
The estimates of these models can be approximated
by using artificial parameters and the linear components,
\begin{equation*}
  \widehat{b}_{i}^{A} \approx \beta_{i}^{A} + s v_{i}^{A}, \quad
  \widehat{b}_{j}^{P} \approx \beta_{j}^{P} - s v_{j}^{P}, \quad
  \widehat{b}_{k}^{C} \approx \beta_{k}^{C} + s v_{k}^{C},
\end{equation*}
taking the medians of the estimates as the particular solutions
and referring to the general solutions (\ref{Eq2:GeneralSolution_APC}).
Here, a small absolute value of $s$ means
the model succeeded in recovering the artificial parameters.
Thus, to evaluate bias, we calculated $s$ such
that the following function satisfies $\mathrm{d}f / \mathrm{d}s = 0$:
\begin{equation*}
  f (s) =
    \sum_{i=1}^{I}
      \left\{ \widehat{b}_{i}^{A} - \left(
          \beta_{i}^{A} + s v_{i}^{A}
        \right)
      \right\}^2 +
      \sum_{j=1}^{J}
      \left\{ \widehat{b}_{j}^{P} - \left(
          \beta_{j}^{P} - s v_{j}^{P}
        \right)
      \right\}^2 +
      \sum_{k=1}^{K}
      \left\{ \widehat{b}_{k}^{C} - \left(
          \beta_{k}^{C} + s v_{k}^{C}
        \right)
      \right\}^2.
\end{equation*}

Table \ref{Tb5:AllResults_BayesianRegularization}
summarizes\footnote{
  The three models of Bayesian regularization
  were implemented using the probabilistic programming language Stan
  \citep{Stan2021}
  and were run in R \citep{R2022}.
  Sampling settings were chains = 4, iter = 6000,
  warmup = 1000, and thin = 5.
  The lower bounds of $\sigma^{A}$, $\sigma^{P}$, and $\sigma^{C}$
  in the random effects model
  were set to 0.05 in order to search for parameters in a wide range,
  as this model can get stuck in locally optimal solutions.
  The parameters of all the models satisfied $\widehat{R}<1.05$.
} the linear components of the artificial parameters
and the evaluation and degree of bias.
The letters A through E that appear in three of the table columns
are used to categorize the results:
A if the absolute value of $s$ is less than 0.02,
B if less than 0.04, C if less than 0.06,
D if less than 0.08, and E if 0.08 or more.
As shown, among the 13 cases of artificial data,
10 cases in the random walk model rated B or better
(i.e., the value of $s$ was less than 0.04)
as compared to 4 cases in the random effects and ridge regression models,
indicating that the random walk model performed relatively well.

\begin{figure}[p]
  \centering
  \includegraphics[height=20.0truecm]
    {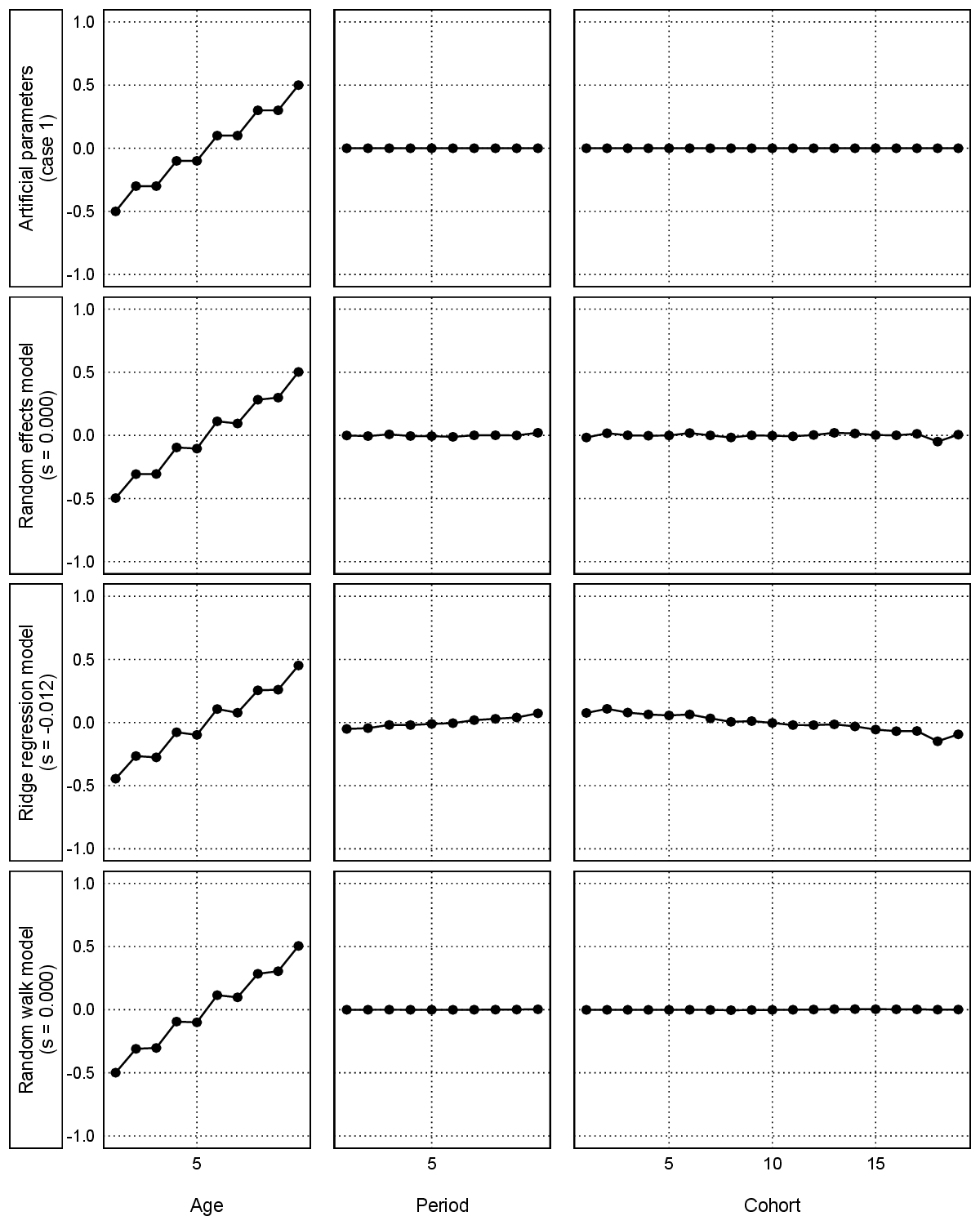}
  \caption{Comparison of the three models' estimates (case 1)}
  \label{Fg5:Results_Case1}
\end{figure}

\begin{figure}[p]
  \centering
  \includegraphics[height=20.0truecm]
    {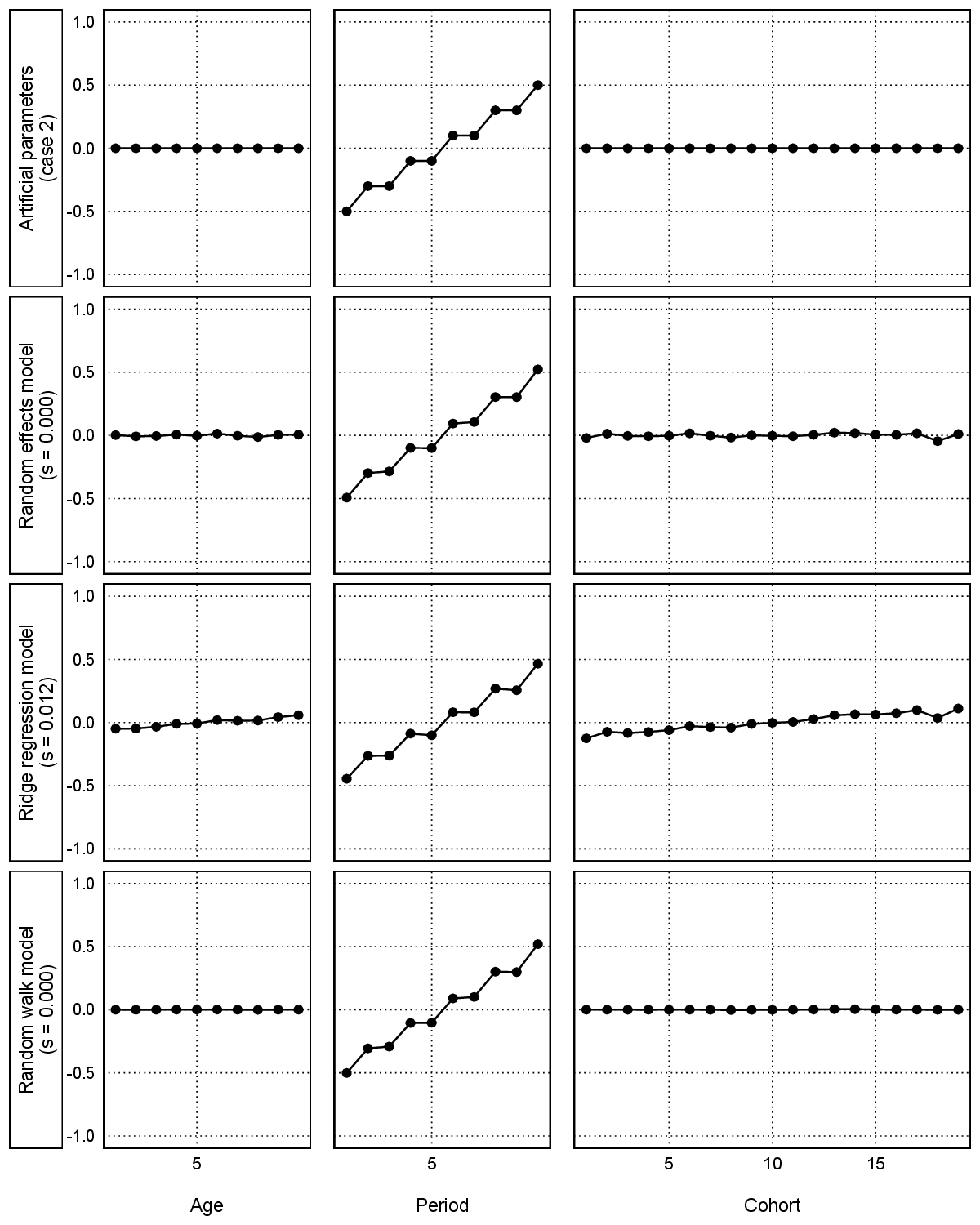}
  \caption{Comparison of the three models' estimates (case 2)}
  \label{Fg5:Results_Case2}
\end{figure}

\begin{figure}[p]
  \centering
  \includegraphics[height=20.0truecm]
    {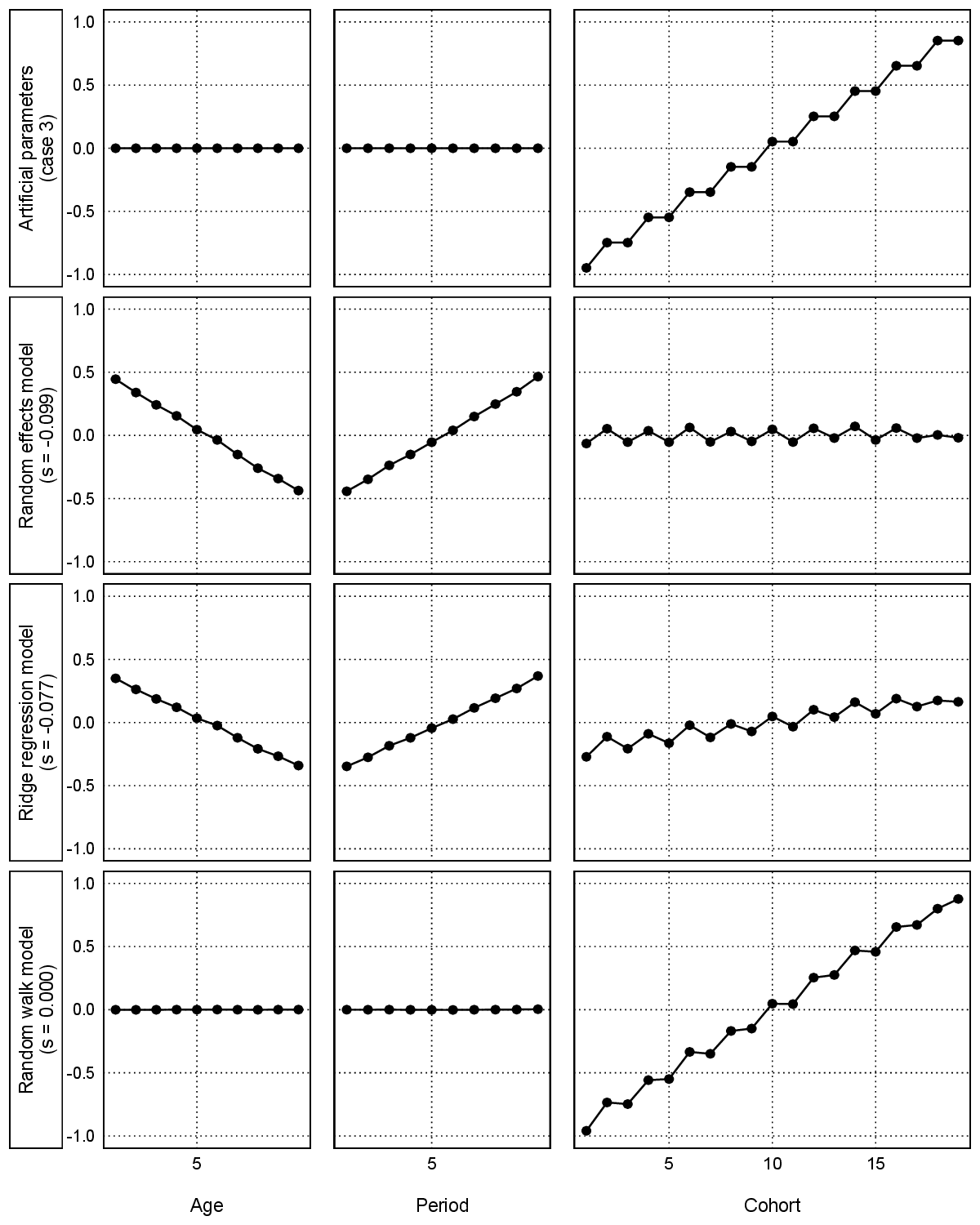}
  \caption{Comparison of the three models' estimates (case 3)}
  \label{Fg5:Results_Case3}
\end{figure}

\begin{figure}[p]
  \centering
  \includegraphics[height=20.0truecm]
    {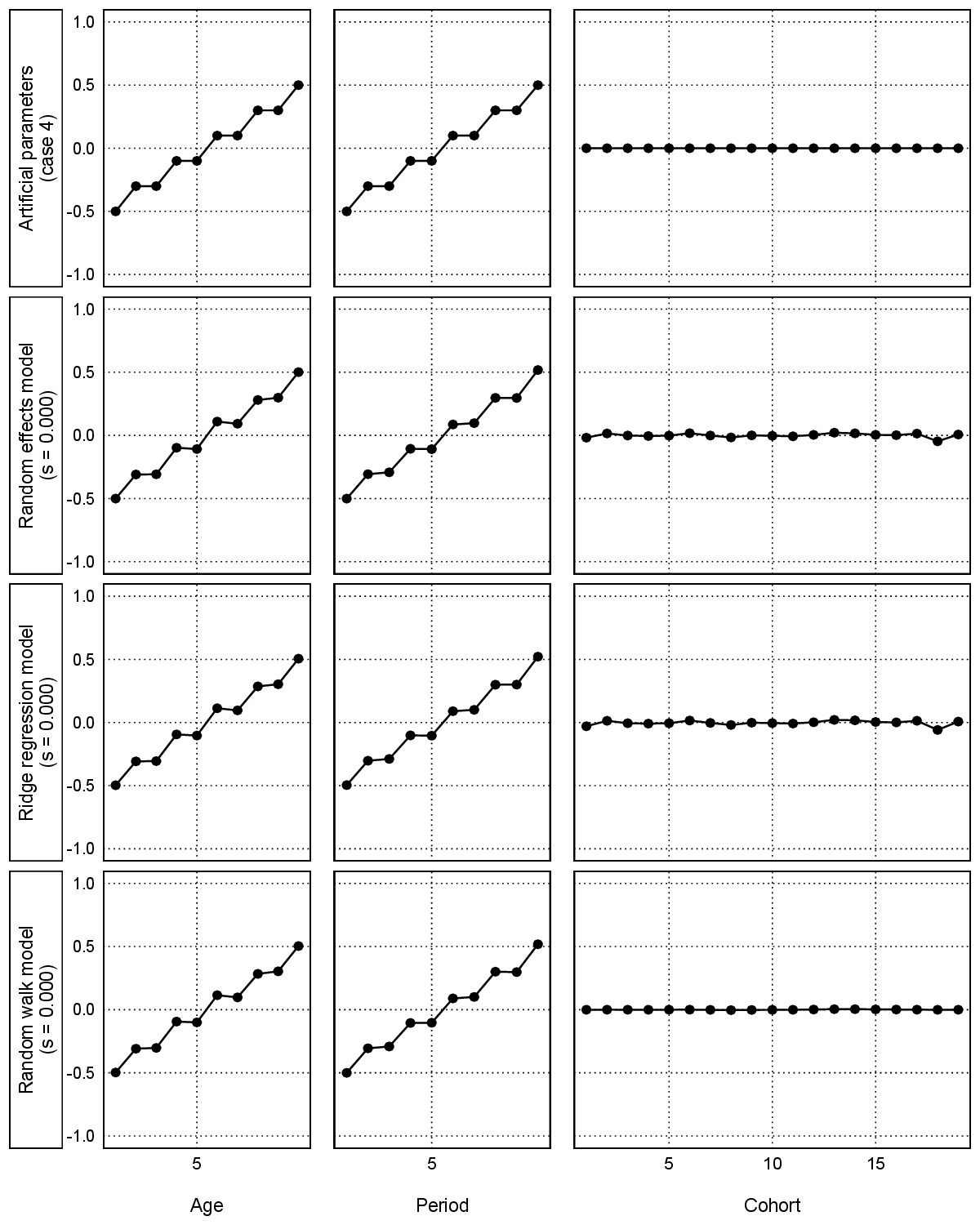}
  \caption{Comparison of the three models' estimates (case 4)}
  \label{Fg5:Results_Case4}
\end{figure}

\begin{figure}[p]
  \centering
  \includegraphics[height=20.0truecm]
    {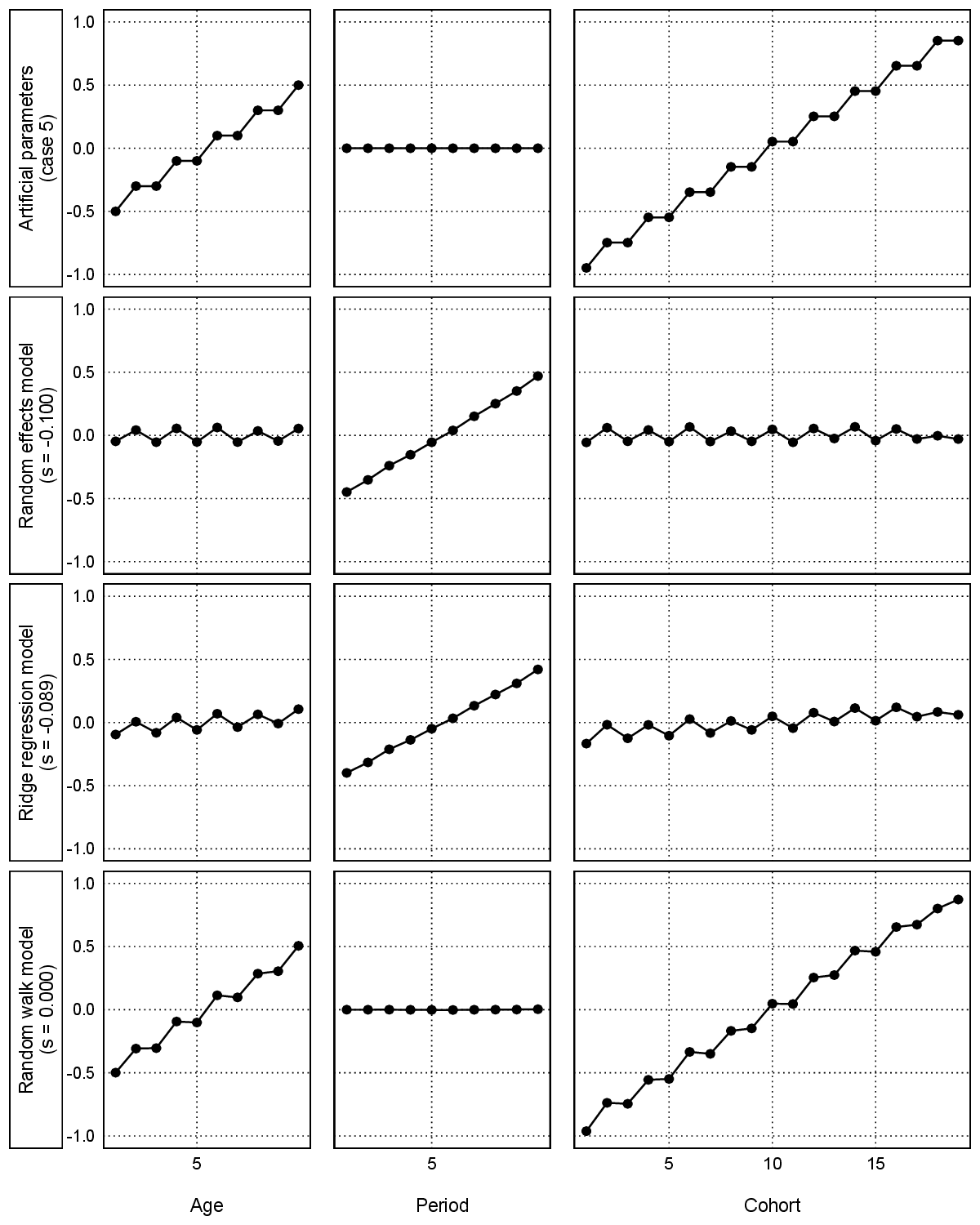}
  \caption{Comparison of the three models' estimates (case 5)}
  \label{Fg5:Results_Case5}
\end{figure}

\begin{figure}[p]
  \centering
  \includegraphics[height=20.0truecm]
    {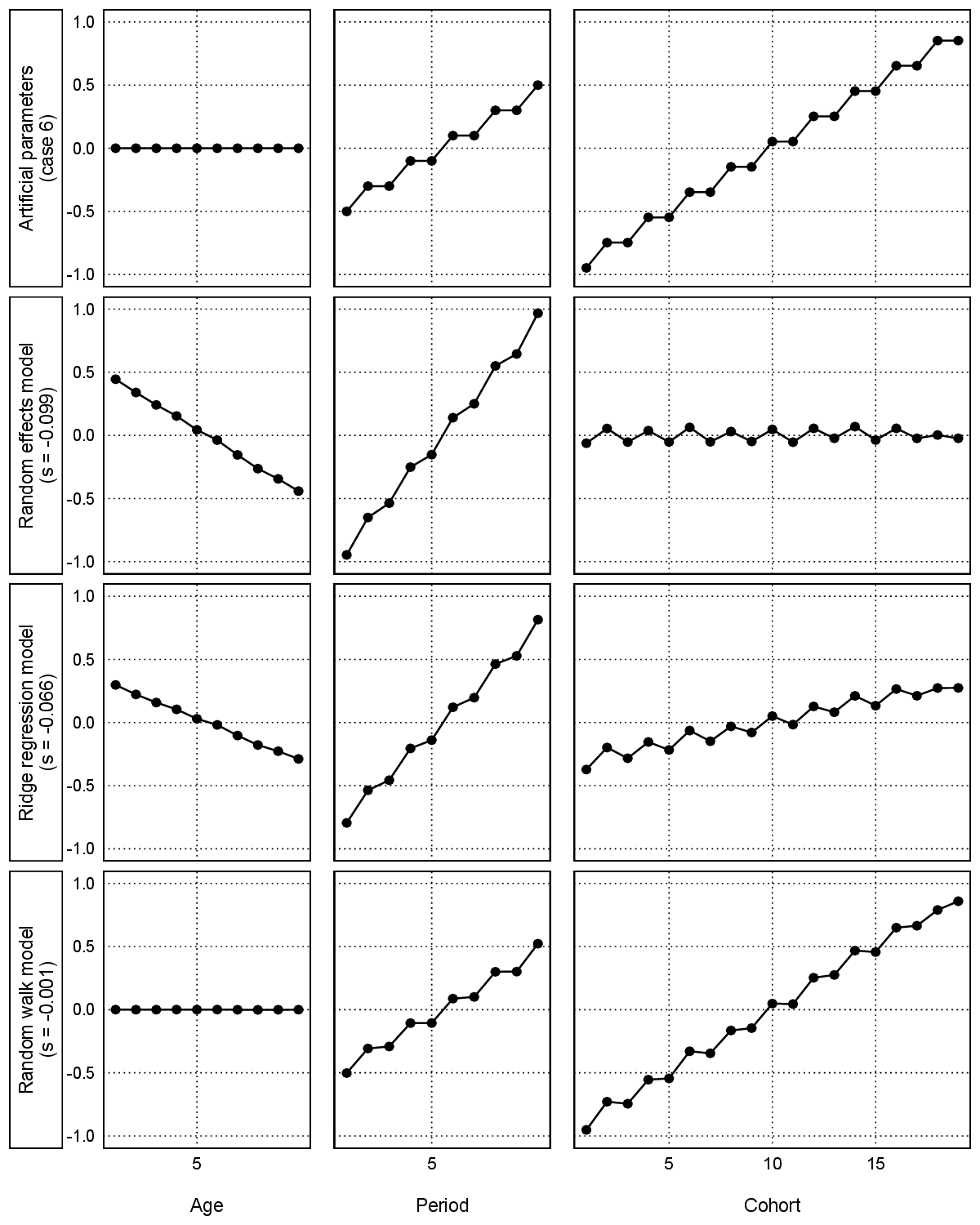}
  \caption{Comparison of the three models' estimates (case 6)}
  \label{Fg5:Results_Case6}
\end{figure}

\begin{figure}[p]
  \centering
  \includegraphics[height=20.0truecm]
    {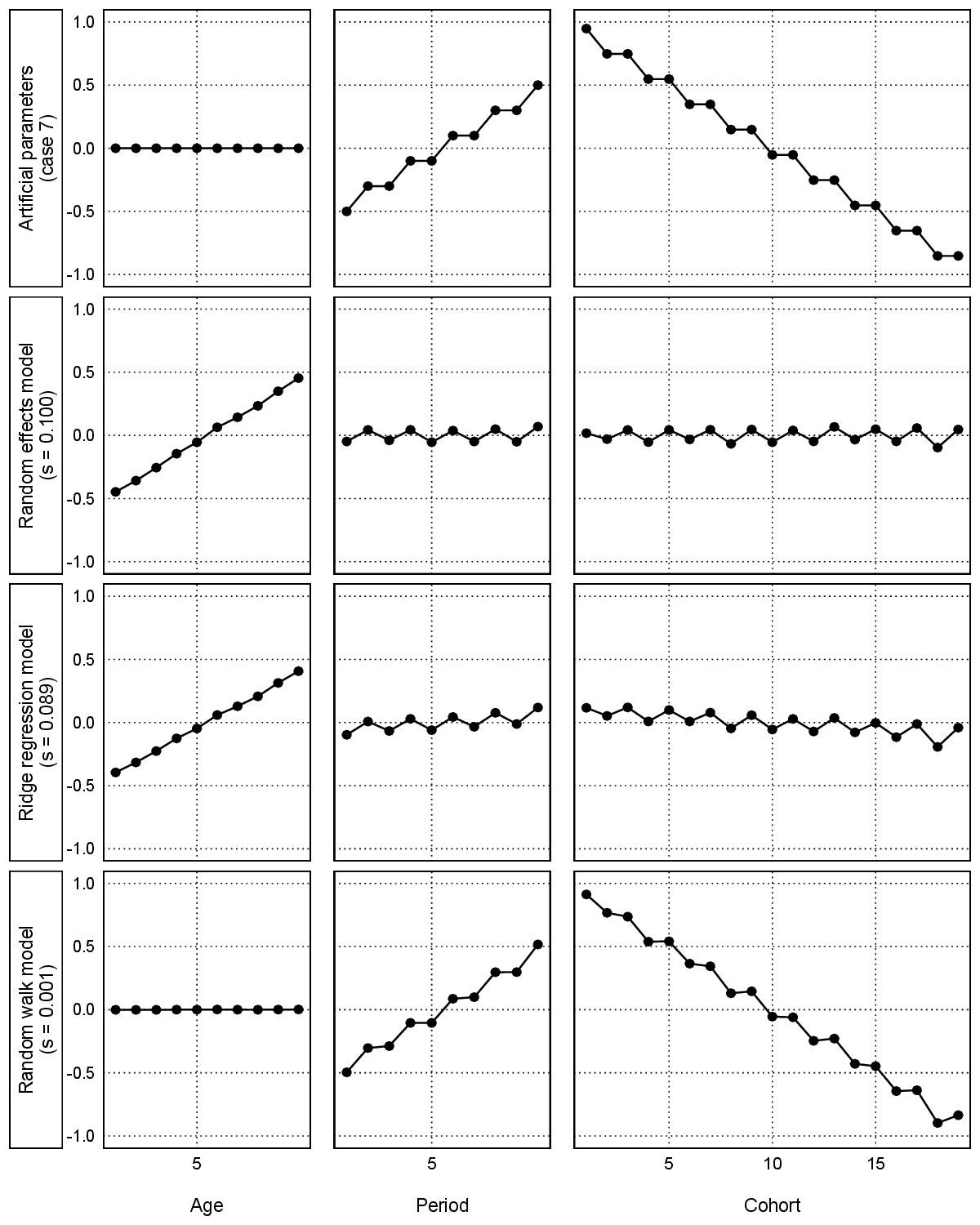}
  \caption{Comparison of the three models' estimates (case 7)}
  \label{Fg5:Results_Case7}
\end{figure}

\begin{figure}[p]
  \centering
  \includegraphics[height=20.0truecm]
    {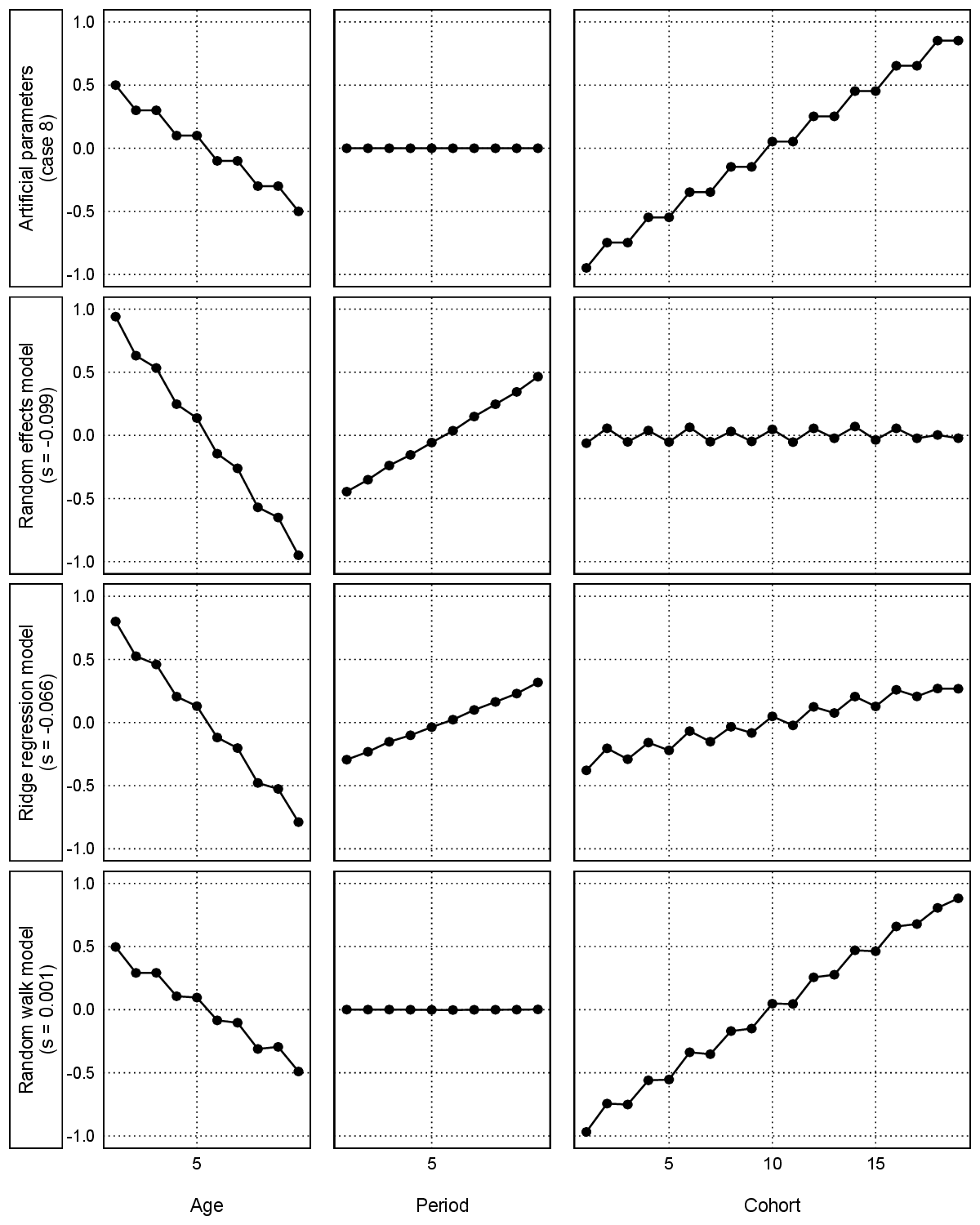}
  \caption{Comparison of the three models' estimates (case 8)}
  \label{Fg5:Results_Case8}
\end{figure}

\begin{figure}[p]
  \centering
  \includegraphics[height=20.0truecm]
    {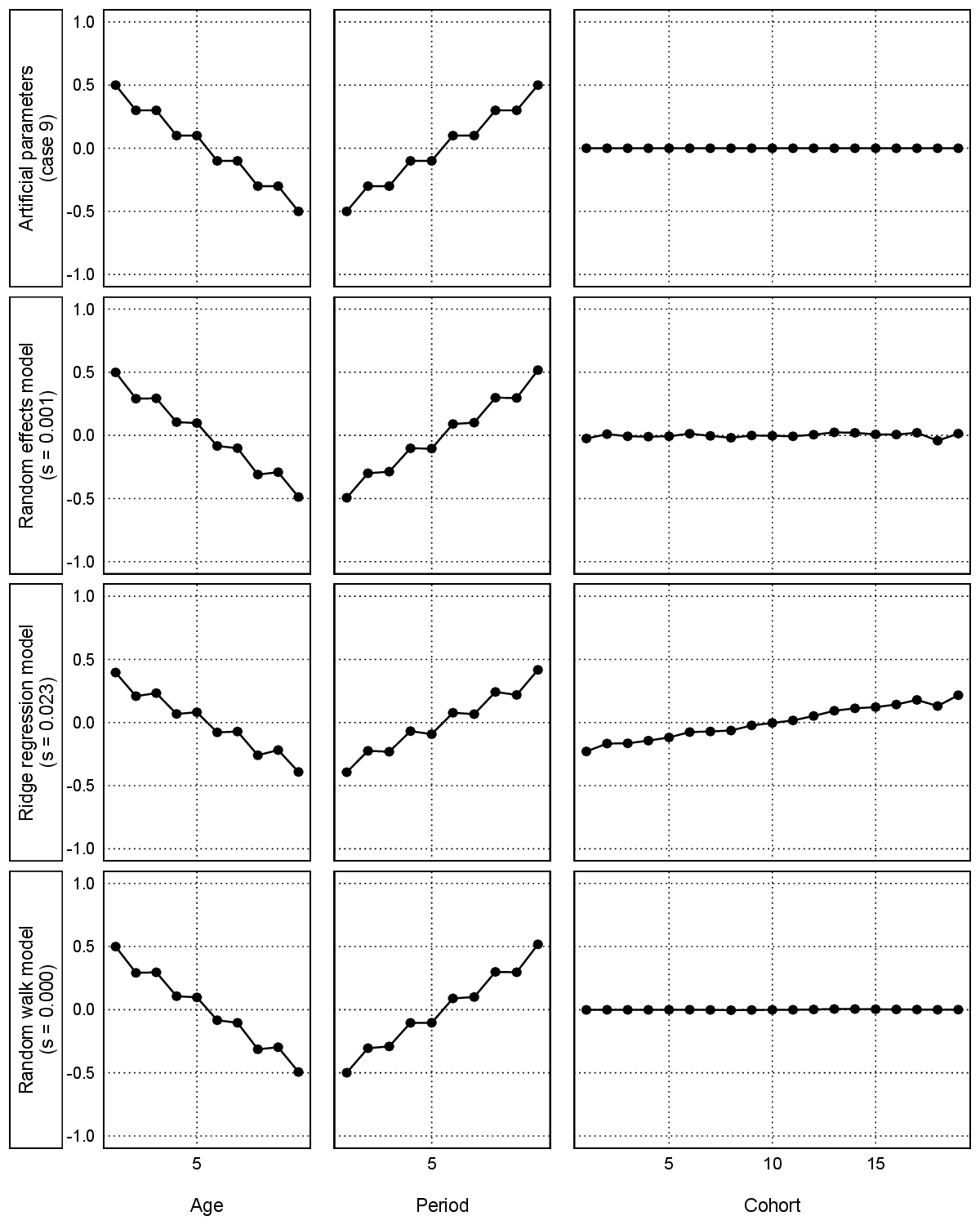}
  \caption{Comparison of the three models' estimates (case 9)}
  \label{Fg5:Results_Case9}
\end{figure}

\begin{figure}[p]
  \centering
  \includegraphics[height=20.0truecm]
    {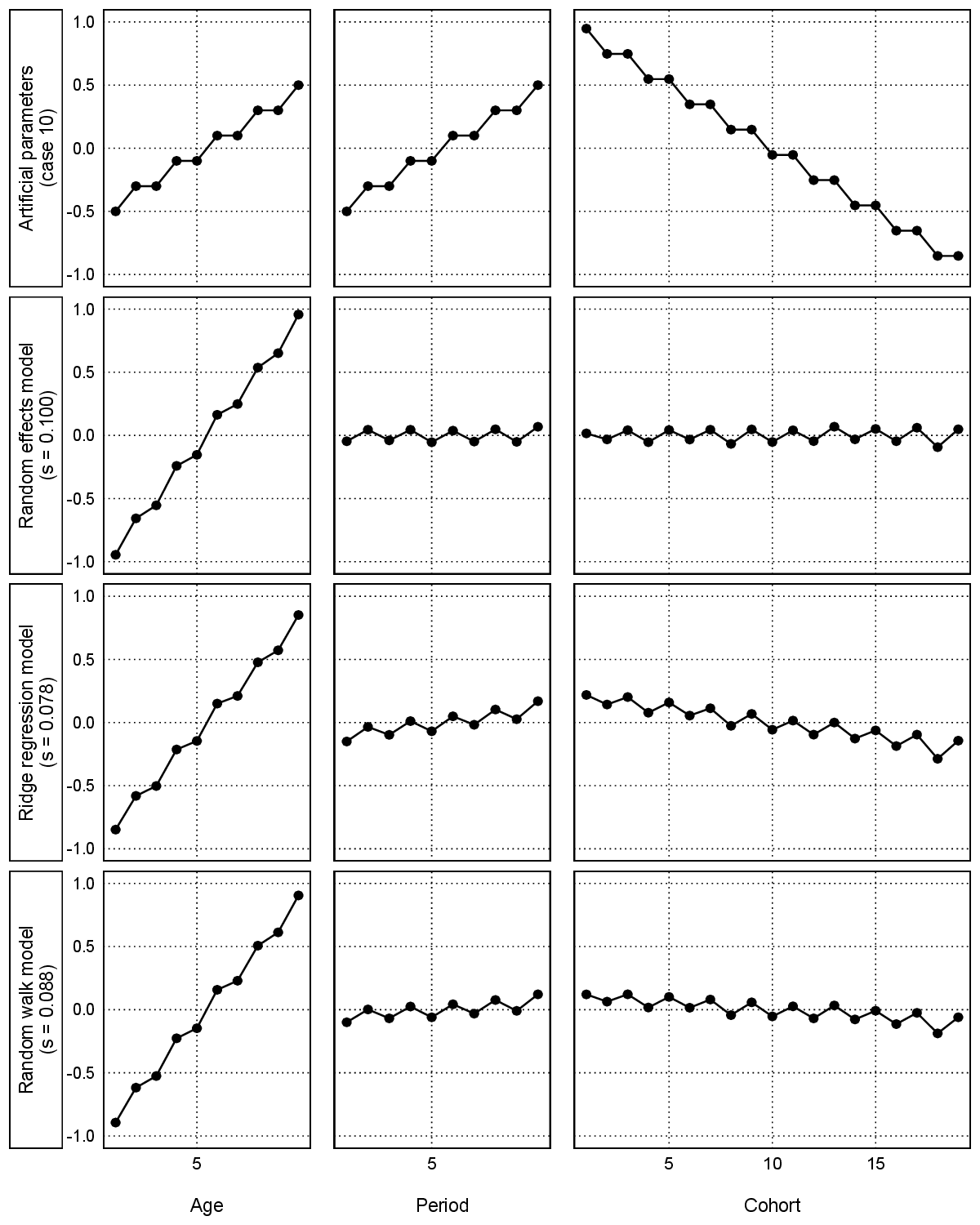}
  \caption{Comparison of the three models' estimates (case 10)}
  \label{Fg5:Results_Case10}
\end{figure}

\begin{figure}[p]
  \centering
  \includegraphics[height=20.0truecm]
    {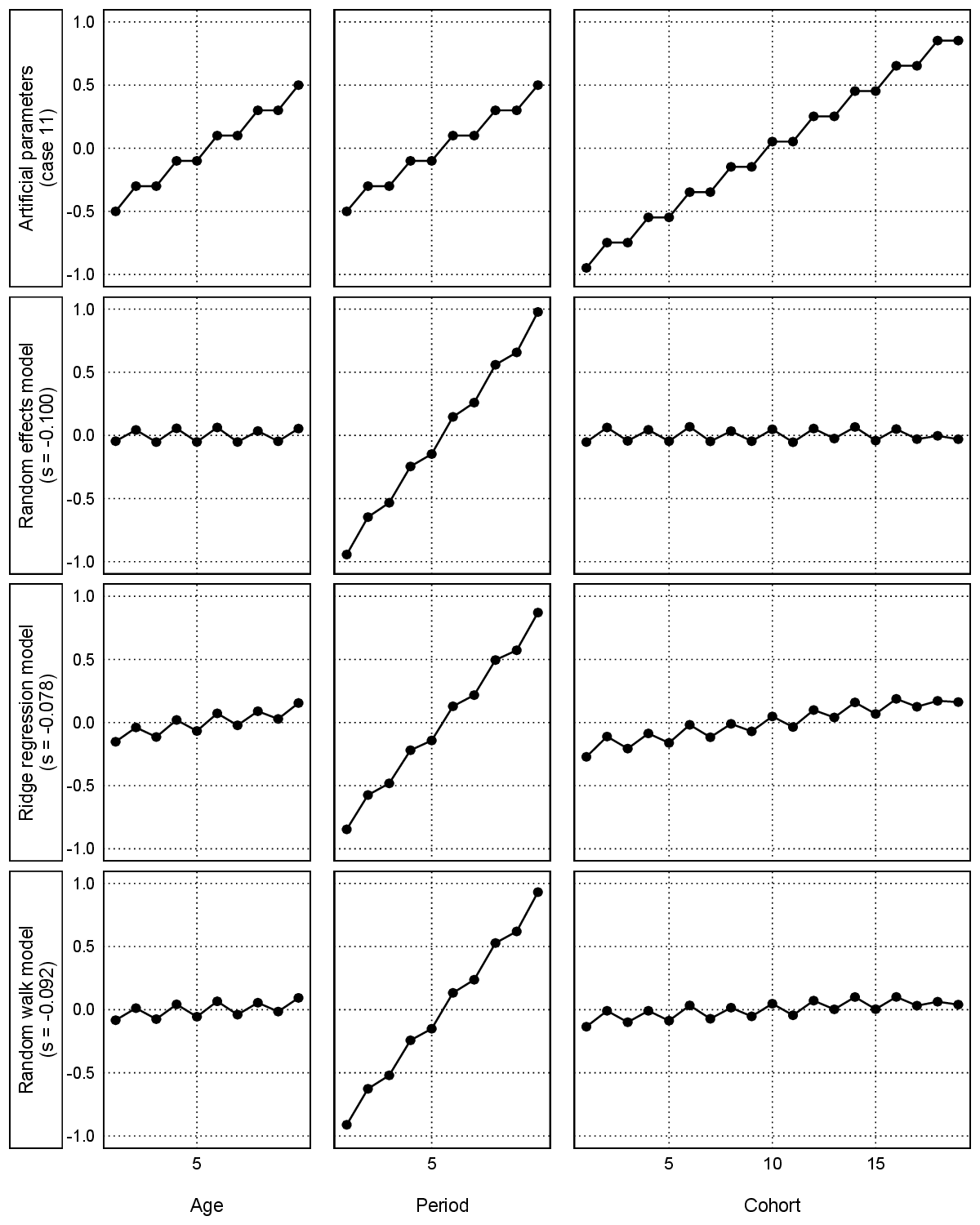}
  \caption{Comparison of the three models' estimates (case 11)}
  \label{Fg5:Results_Case11}
\end{figure}

\begin{figure}[p]
  \centering
  \includegraphics[height=20.0truecm]
    {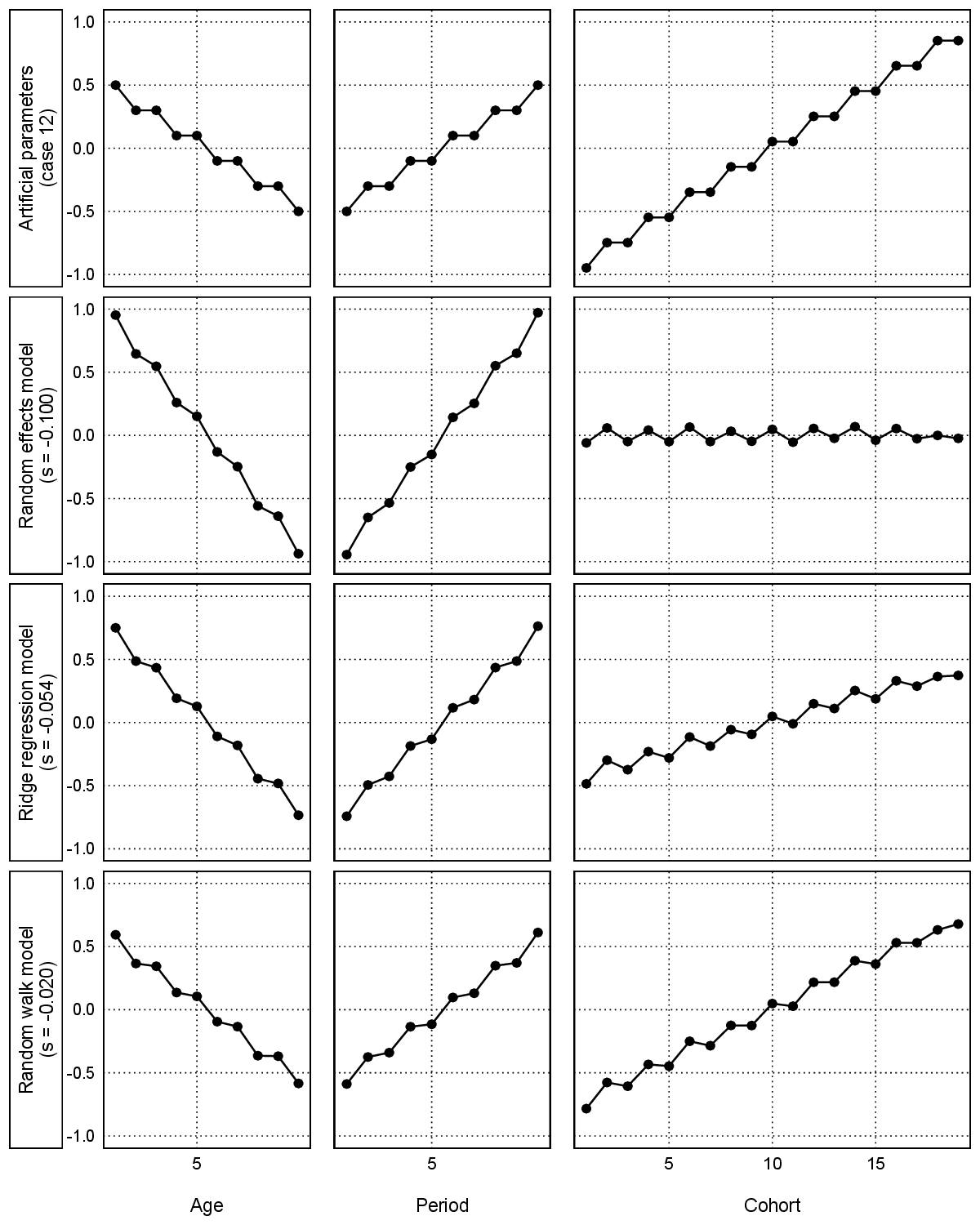}
  \caption{Comparison of the three models' estimates (case 12)}
  \label{Fg5:Results_Case12}
\end{figure}

\begin{figure}[p]
  \centering
  \includegraphics[height=20.0truecm]
    {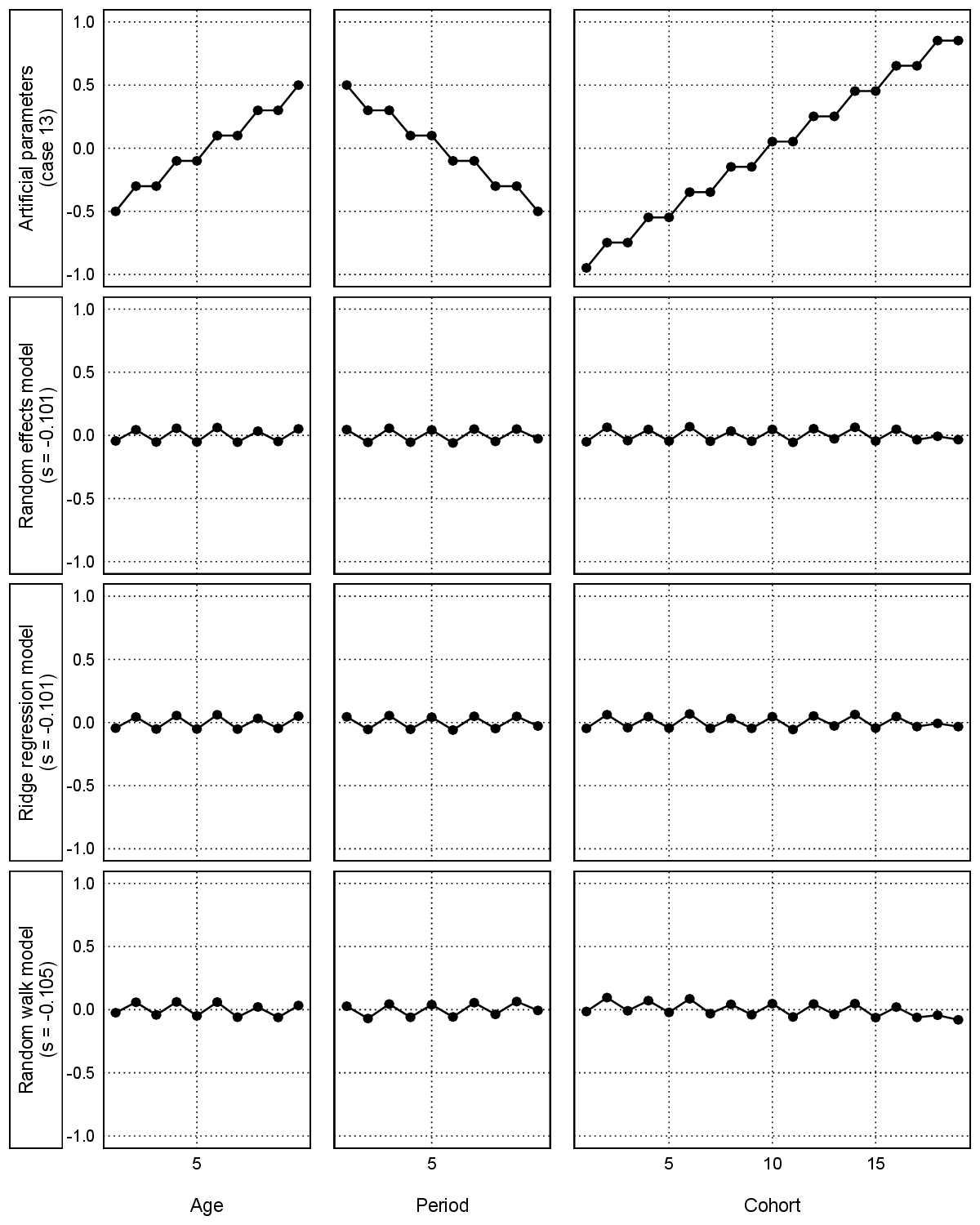}
  \caption{Comparison of the three models' estimates (case 13)}
  \label{Fg5:Results_Case13}
\end{figure}

\clearpage

The cases in which the models failed to effectively recover
the artificial parameters each contain the linear component
of the cohort effects.
First, constraints shrinking the parameters,
such as in Bayesian regularization, always fail case 13,
where the linear components completely cancel.
Moreover, we found
the estimated linear component of the cohort effects to be close to zero,
because $(\mathrm{C}) \, +$ leads to $s<0$ and
$(\mathrm{C}) \, -$ leads to $s>0$.
Specifically, Figure \ref{Fg5:Results_Case3},
which visualizes case 3, shows
that the estimated slope of the cohort effects becomes horizontal,
and the linear component is incorrectly assigned to the other effects.
The random effects and ridge regression models obtain the estimates
like the artificial parameters in case 9 (Figure \ref{Fg5:Results_Case9})
because the age effects have a negative slope
and the period effects have a positive slope.
However, the random walk model did not underestimate
the linear component of the cohort effects.

\section{Conclusion}

This paper reviewed three models of Bayesian regularization
applying normal distributions in APC analysis:
the random effects model, the ridge regression model,
and the random walk model.
In addition, by expressing the linear components of the three effects
using centering indexes,
we confirmed the mathematical mechanism of bias for the three models.
Then, this paper proposed a systematic simulation
that generated artificial data
with a combination of linear components
to evaluate the performance of the three models
and demonstrated that the random walk model effectively recovered
the artificial parameters.

Our study attempted to statistically estimate
the mathematically indistinguishable linear components
by using the mathematically identifiable nonlinear components
and log prior probabilities
in order to overcome the well-known identification problem
in APC analysis.
We then assumed that factors containing the linear components
also have nonlinear components.
Here, the models of Bayesian regularization have two main biases:
(1) it always fails to estimate the cases
where the linear components completely cancel each other, and
(2) the regularization does not work uniformly.
Specifically, the index weights have a large influence
on the cohort effects owing to $K=I+J-1$
and this constraint drives the linear component of the cohort effects
close to zero.

According to the systematic simulation used in the study,
the random effects and ridge regression models performed poorly.
In particular, the random effects model\footnote{
  This paper does not recommend using proxy variables
  without identifying the three effects,
  although some studies discuss the random effects model
  that includes them.
  Putting proxy variables of period in the model
  may yield erroneous results. In case 3, for example,
  there are no period effects,
yet the random effects model still estimates
  the linear components of the period effects.
}
reproduces the findings of previous studies
that the linear component of the cohort effects becomes flat.
The ridge regression model equivalent to the intrinsic estimator
gives results that are comparable to the mean of
all the particular solutions,
which seems statistically desirable.
However, this paper evaluates the ridge regression model as problematic,
as Table \ref{Tb5:AllResults_BayesianRegularization} shows
that the model not only underestimates
the linear component of the cohort effects
but also overestimates that component in  case 9
(Figure \ref{Fg5:Results_Case9}).

By contrast, the random walk model, unlike the other two models,
mitigates underestimating the linear component of the cohort effects
caused by the index weights.
The simulations in this paper show
that the random walk model performs well
even when the artificial parameters are generated
by trigonometric functions rather than random walks.
However, it should not be forgotten that
this model was unable to produce useful estimates\footnote{
  The performance of the random walk model may be improved,
  as in cases 1 to 9,
  if the period or cohort effects do not continue to change linearly
  as the observation period becomes longer in the real data.
} in cases 10, 11, and 13, as shown
in Table \ref{Tb5:AllResults_BayesianRegularization}.

Finally, although this paper simulated artificial data
focusing on combinations of the linear components,
we did not verify various patterns of the nonlinear components.
If the absolute values of the nonlinear components are small,
there are few clues to separate the linear components.
On the other hand,
when the absolute value of $\beta^{C[\mathit{NL}]}$ is large,
the random effects model may recover
the linear component by increasing $\sigma^{C}$.
Therefore, we hope other researchers verify the models of APC analysis
through the systematic simulation in this paper
that allows to arbitrarily set the amount of change in the artificial parameters.
In addition, since Bayesian regularization is influenced by index weights,
future studies need to investigate
whether tuning $\sigma^{A}$, $\sigma^{P}$, and $\sigma^{C}$
by considering $v^{A}$, $v^{P}$, and $v^{C}$ is effective.

\clearpage

\bibliographystyle{apalike}
\bibliography{References}

\begin{thebibliography}{}

\bibitem[Fosse and Winship, 2019]{Fosse2019}
Fosse, E. and Winship, C. (2019).
\newblock Analyzing age-period-cohort data: A review and critique.
\newblock {\em Annual Review of Sociology}, 45(1):467--492.
\newblock doi:10.1146/annurev-soc-073018-022616.

\bibitem[Fu, 2008]{Fu2008}
Fu, W.~J. (2008).
\newblock A smoothing cohort model in age-period-cohort analysis with
  applications to homicide arrest rates and lung cancer mortality rates.
\newblock {\em Sociological Methods \& Research}, 36(3):327--361.
\newblock doi:10.1177/0049124107310637.

\bibitem[Kupper et~al., 1985]{Kupper1985}
Kupper, L.~L., Janis, J.~M., Karmous, A., and Greenberg, B.~G. (1985).
\newblock Statistical age-period-cohort analysis: a review and critique.
\newblock {\em Journal of Chronic Disease}, 38(10):811--830.

\bibitem[Nakamura, 1986]{Nakamura1986}
Nakamura, T. (1986).
\newblock Bayesian cohort models for general cohort table analyses.
\newblock {\em Annals of the Institute of Statistical Mathematics},
  38(B):353--370.

\bibitem[O'Brien, 2014]{O'Brien2014}
O'Brien, R. (2014).
\newblock {\em Age-Period-Cohort Models: Approaches and Analyses with Aggregate
  Data}.
\newblock CRC, Florida, 1st edition.
\newblock doi:10.1201/b17286.

\bibitem[{R Core Team}, 2022]{R2022}
{R Core Team} (2022).
\newblock {\em R: A Language and Environment for Statistical Computing}.
\newblock R Foundation for Statistical Computing, Vienna, Austria.
\newblock version 4.1.3. https://www.R-project.org/.

\bibitem[Ryder, 1965]{Ryder1965}
Ryder, N.~B. (1965).
\newblock The cohort as a concept in the study of social change.
\newblock {\em American Sociological Review}, 30(6):843--861.
\newblock doi:10.2307/2090964.

\bibitem[Sakaguchi and Nakamura, 2019]{Sakaguchi2019}
Sakaguchi, N. and Nakamura, T. (2019).
\newblock Age-period-cohort (apc) model as the mixed effects model: Comparison
  of the hierarchical apc model and the bayesian apc model.
\newblock {\em Sociological Theory and Methods}, 34(1).
\newblock (in Japanese).

\bibitem[Schmid and Held, 2007]{Schmid2007}
Schmid, V.~J. and Held, L. (2007).
\newblock Bayesian age-period-cohort modeling and prediction: Bamp.
\newblock {\em Journal of Statistical Software}, 21(8):1--15.
\newblock doi:10.18637/jss.v021.i08.

\bibitem[{Stan Development Team}, 2021]{Stan2021}
{Stan Development Team} (2021).
\newblock {RStan}: the {R} interface to {Stan}.
\newblock R package version 2.21.3. https://mc-stan.org/.

\bibitem[Yang, 2006]{Yang2006}
Yang, Y. (2006).
\newblock Bayesian inference for hierarchical age-period-cohort models of
  repeated cross-section survey data.
\newblock {\em Sociological Methodology}, 36(1):39--74.
\newblock doi:10.\allowbreak{}1111/j.1467-9531.\allowbreak{}2006.00174.x.

\bibitem[Yang et~al., 2004]{Yang2004}
Yang, Y., Fu, W.~J., and Land, K.~C. (2004).
\newblock A methodological comparison of age-period-cohort models: The
  intrinsic estimator and conventional generalized linear models.
\newblock {\em Sociological Methodology}, 34(1):75--110.
\newblock doi:10.1111/j.0081-1750.2004.00148.x.

\bibitem[Yang and Land, 2013]{Yang2013}
Yang, Y. and Land, K.~C. (2013).
\newblock {\em Age-Period-Cohort Analysis: New Models, Methods, and Empirical
  Applications}.
\newblock CRC, Florida.

\end{thebibliography}

\clearpage

\appendix

\section*{Appendix A. Stan codes}

Appendix A describes the Stan codes for Bayesian regularization
using normal distributions.

\leavevmode \\
\noindent
{\textbf{Random effects model}} \quad (\texttt{RE\_Normal\_APC.stan})
\begin{lstlisting}
data {
  int N;
  int I;
  int J;
  int K;
  int L;
  matrix[N, L]  X;
  vector[N]     Y;
  real<lower=0> Min;
}

parameters {
  real<lower=0> raw_A;
  real<lower=0> raw_P;
  real<lower=0> raw_C;
  vector[I]     std_A;
  vector[J]     std_P;
  vector[K]     std_C;
  real<lower=0> sigma;
  real          b_0;
}

transformed parameters {
  vector[I]     b_A;
  vector[J]     b_P;
  vector[K]     b_C;
  vector[I+J]   b_AP;
  vector[L]     b;
  real<lower=0> sigma_A;
  real<lower=0> sigma_P;
  real<lower=0> sigma_C;

  sigma_A = raw_A + Min;
  sigma_P = raw_P + Min;
  sigma_C = raw_C + Min;

  b_A = sigma_A * std_A;
  b_P = sigma_P * std_P;
  b_C = sigma_C * std_C;

  b_AP = append_row(b_A,  b_P);
  b    = append_row(b_AP, b_C);
}

model {
  target += std_normal_lpdf(std_A);
  target += std_normal_lpdf(std_P);
  target += std_normal_lpdf(std_C);

  target += normal_id_glm_lpdf(Y | X, b_0, b, sigma);
}
\end{lstlisting}

\newpage
\indent
For efficient sampling,
this paper reparameterizes
the random effects model by the following procedure:
we generate standard distributed parameters,
\begin{align*}
  \mathit{std}_{i}^{A}  & \sim \mathrm{Normal} \,
    (0, 1 ) & i &  = 1,
    \ldots , I, \\
  \mathit{std}_{j}^{P} & \sim \mathrm{Normal} \,
    (0, 1 ) & j & = 1,
    \ldots , J,  \\
  \mathit{std}_{k}^{C} & \sim \mathrm{Normal} \,
    (0, 1 ) & k &  = 1,
    \ldots , K,
\end{align*}
and create the three effects,
\begin{equation*}
  b_{i}^{A} = \sigma^{A} \cdot \mathit{std}_{i}^{A}, \quad
  b_{j}^{P} = \sigma^{P} \cdot \mathit{std}_{j}^{P}, \quad
  b_{k}^{C} = \sigma^{C} \cdot \mathit{std}_{k}^{C}.
\end{equation*}

\leavevmode \\
\noindent
{\textbf{Ridge regression model}} \quad (\texttt{RR\_Normal\_APC.stan})
\begin{lstlisting}
data {
  int N;
  int I;
  int J;
  int K;
  int L;
  matrix[N, L] X;
  vector[N]    Y;
}

parameters {
  vector[I]     std_A;
  vector[J]     std_P;
  vector[K]     std_C;
  real<lower=0> sigma;
  real<lower=0> lambda;
  real          b_0;
}

transformed parameters {
  vector[I]   b_A;
  vector[J]   b_P;
  vector[K]   b_C;
  vector[I+J] b_AP;
  vector[L]   b;

  b_A = lambda * std_A;
  b_P = lambda * std_P;
  b_C = lambda * std_C;

  b_AP = append_row(b_A,  b_P);
  b    = append_row(b_AP, b_C);
}

model {
  target += std_normal_lpdf(std_A);
  target += std_normal_lpdf(std_P);
  target += std_normal_lpdf(std_C);

  target += normal_id_glm_lpdf(Y | X, b_0, b, sigma);
}
\end{lstlisting}

We reparameterize the ridge regression model
by unifying the standard deviations of the priors,
referring to the procedure for random effects models,
\begin{equation*}
  b_{i}^{A} = \lambda \cdot \mathit{std}_{i}^{A}, \quad
  b_{j}^{P} = \lambda \cdot \mathit{std}_{j}^{P}, \quad
  b_{k}^{C} = \lambda \cdot \mathit{std}_{k}^{C}.
\end{equation*}

\newpage

\noindent
{\textbf{Random walk model}} \quad (\texttt{RW\_Normal\_APC.stan})
\begin{lstlisting}
data {
  int N;
  int I;
  int J;
  int K;
  int L;
  matrix[N, L] X;
  vector[N]    Y;
}

parameters {
  vector[I-1]   std_A;
  vector[J-1]   std_P;
  vector[K-1]   std_C;
  real<lower=0> sigma;
  real<lower=0> sigma_A;
  real<lower=0> sigma_P;
  real<lower=0> sigma_C;
  real          b_0;
}

transformed parameters {
  vector[I-1] d_A;
  vector[I-1] tmp_d_A;
  vector[I]   b_A;
  vector[J-1] d_P;
  vector[J-1] tmp_d_P;
  vector[J]   b_P;
  vector[K-1] d_C;
  vector[K-1] tmp_d_C;
  vector[K]   b_C;
  vector[I+J] b_AP;
  vector[L]   b;

  // Age
  d_A = sigma_A * std_A;

  for (a in 1:(I-1)) {
    tmp_d_A[a] = (I-a) * d_A[a];
  }

  b_A[1] = -(1.0/I) * sum(tmp_d_A);

  for (i in 2:I) {
    b_A[i] = b_A[1] + sum(d_A[1:(i-1)]);
  }

  // Period
  d_P = sigma_P * std_P;

  for (p in 1:(J-1)) {
    tmp_d_P[p] = (J-p) * d_P[p];
  }

  b_P[1] = -(1.0/J) * sum(tmp_d_P);

  for (j in 2:J) {
    b_P[j] = b_P[1] + sum(d_P[1:(j-1)]);
  }

  // Cohort
  d_C = sigma_C * std_C;

  for (c in 1:(K-1)) {
    tmp_d_C[c] = (K-c) * d_C[c];
  }

  b_C[1] = -(1.0/K) * sum(tmp_d_C);

  for (k in 2:K) {
    b_C[k] = b_C[1] + sum(d_C[1:(k-1)]);
  }


  b_AP = append_row(b_A,  b_P);
  b    = append_row(b_AP, b_C);
}

model {
  target += std_normal_lpdf(std_A);
  target += std_normal_lpdf(std_P);
  target += std_normal_lpdf(std_C);

  target += normal_id_glm_lpdf(Y | X, b_0, b, sigma);
}
\end{lstlisting}

Before implementing reparameterization,
we discuss the difference terms of the adjacent parameters,
\begin{equation*}
  d_{i}^{A} = b_{i+1}^{A} - b_{i}^{A}, \quad
  d_{j}^{P} = b_{j+1}^{P} - b_{j}^{P}, \quad
  d_{k}^{C} = b_{k+1}^{C} - b_{k}^{C}.
\end{equation*}
Here, we can rewrite the random walk assumptions by difference terms,
\begin{align*}
  d_{i}^{A}  & \sim \mathrm{Normal} \,
    (0, \sigma^{A} ) & i &  = 1,
    \ldots , I-1, \\
  d_{j}^{P} & \sim \mathrm{Normal} \,
    (0, \, \sigma^{P} ) & j & = 1,
    \ldots , J-1,  \\
  d_{k}^{C} & \sim \mathrm{Normal} \,
    (0, \sigma^{C} ) & k &  = 1,
    \ldots , K-1.
\end{align*}
The original parameters can be calculated
by the first terms and the difference terms,
\begin{equation*}
  b_{i}^{A} = b_{1}^{A} + \sum_{a=1}^{i-1} d_{a}^{A}, \quad
  b_{j}^{P} = b_{1}^{P} + \sum_{p=1}^{j-1} d_{p}^{P}, \quad
  b_{k}^{C} = b_{1}^{C} + \sum_{c=1}^{k-1} d_{c}^{C},
\end{equation*}
and since each effect satisfies the sum-to-zero condition,
we obtain the following relations,
\begin{align*}
  \sum_{i=1}^{I} b_{i}^{A}
     & = I \cdot b_{1}^{A} +
     \sum_{a=1}^{I-1} \, (I-a) \, d_{a}^{A}
     = 0, \\
  \sum_{j=1}^{J} b_{j}^{P}
     & = J \cdot b_{1}^{P} +
     \sum_{p=1}^{J-1} \, (J-p) \, d_{p}^{P}
     = 0, \\
  \sum_{k=1}^{K} b_{k}^{C}
     & = K \cdot b_{1}^{C} +
     \sum_{c=1}^{K-1} \, (K-c) \, d_{c}^{C}
     = 0,
\end{align*}
and derive the conditions that the first terms need to satisfy
\begin{equation*}
  b_{1}^{A} = - \frac{1}{I} \sum_{a=1}^{I-1} \, (I-a) \, d_{a}^{A}, \quad
  b_{1}^{P} = - \frac{1}{J} \sum_{p=1}^{J-1} \, (J-p) \, d_{p}^{P}, \quad
  b_{1}^{C} = - \frac{1}{K} \sum_{c=1}^{K-1} \, (K-c) \, d_{c}^{C}.
\end{equation*}

Thus, this paper reparameterizes the random walk model
by the following procedure:
(1)~we generate the standard distributed parameters,
\begin{align*}
  \mathit{std}_{i}^{A}  & \sim \mathrm{Normal} \,
    (0, 1 ) & i &  = 1,
    \ldots , I-1, \\
  \mathit{std}_{j}^{P} & \sim \mathrm{Normal} \,
    (0, 1 ) & j & = 1,
    \ldots , J-1,  \\
  \mathit{std}_{k}^{C} & \sim \mathrm{Normal} \,
    (0, 1 ) & k &  = 1,
    \ldots , K-1,
\end{align*}
and create the difference terms,
\begin{equation*}
  d_{i}^{A} = \sigma^{A} \cdot \mathit{std}_{i}^{A}, \quad
  d_{j}^{P} = \sigma^{P} \cdot \mathit{std}_{j}^{P}, \quad
  d_{k}^{C} = \sigma^{C} \cdot \mathit{std}_{k}^{C},
\end{equation*}
and (2) compute the three effects by the difference terms,
\begin{align*}
  b_{1}^{A} & = - \frac{1}{I} \sum_{a=1}^{I-1} \, (I-a) \, d_{a}^{A}, & &
    & b_{i}^{A} = b_{1}^{A} + \sum_{a=1}^{i-1} d_{a}^{A} & & \qquad
    & i=1, \ldots, I-1, \\
  b_{1}^{P} & = - \frac{1}{J} \sum_{p=1}^{J-1} \, (J-p) \, d_{p}^{P}, & &
    & b_{j}^{P} = b_{1}^{P} + \sum_{p=1}^{j-1} d_{p}^{P} & & \qquad
    & j=1, \ldots, J-1, \\
  b_{1}^{C} & = - \frac{1}{K} \sum_{c=1}^{K-1} \, (K-c) \, d_{c}^{C}, & &
    & b_{k}^{C} = b_{1}^{C} + \sum_{c=1}^{k-1} d_{c}^{C} & & \qquad
    & k=1, \ldots, K-1.
\end{align*}


\section*{Appendix B. R codes}

Appendix B describes
the R codes generate artificial data for case 8
and execute three models of Bayesian regularization.
Before running the simulation,
it is necessary to specify the directory
where the Stan files are located.
\begin{lstlisting}
###### Set artificial parameters ######------------

library(rstan)


# case 8

I <- 10;  J <- 10;  K <- I+J-1;  L <- I+J+K
Times <- 10;  N <- I*J*Times;  gamma <- 0.1

beta_A_L <- -0.1;  beta_A_NL <- -0.05
beta_P_L <-  0.0;  beta_P_NL <-  0.00
beta_C_L <-  0.1;  beta_C_NL <-  0.05



###### Generate artificial data  ######------------

beta_A <- numeric(I);  beta_P <- numeric(J);  beta_C <- numeric(K)

for (i in 1:I) {
  beta_A[i] <- - (beta_A_NL / (2 * I)) * (cos(pi * I) -1) +
    beta_A_L * (i - (I+1)/2) + beta_A_NL * cos(pi * i)
}

for (j in 1:J) {
  beta_P[j] <- - (beta_P_NL / (2 * J)) * (cos(pi * J) -1) +
    beta_P_L * (j - (J+1)/2) + beta_P_NL * cos(pi * j)
}

for (k in 1:K) {
  beta_C[k] <- - (beta_C_NL / (2 * K)) * (cos(pi * K) -1) +
    beta_C_L * (k - (K+1)/2) + beta_C_NL * cos(pi * k)
}



Index_A <- rep(rep(1:I, times = J), times = Times)
Index_P <- rep(rep(1:J, each  = I), times = Times)
Index_C <- Index_P - Index_A + I  # k = j-i+I



X_A <- numeric();  X_P <- numeric();  X_C <- numeric()

for (i in 1:I) {
  x_A <- ifelse(Index_A == i, 1, 0);  X_A <- cbind(X_A, x_A)
}

for (j in 1:J) {
  x_P <- ifelse(Index_P == j, 1, 0);  X_P <- cbind(X_P, x_P)
}

for (k in 1:K) {
  x_C <- ifelse(Index_C == k, 1, 0);  X_C <- cbind(X_C, x_C)
}



X <- cbind(X_A, X_P, X_C)  # Design matrix
beta <- c(beta_A, beta_P, beta_C)

set.seed(1234)
Y <- as.vector(X %*% beta + rnorm(N, 0, gamma))



###### Execute Bayesian APC analysis ######------------

# Random effects model

StanData <- list(Y=Y, X=X, N=N, I=I, J=J, K=K, L=L, Min=0.05)

Pars <- c("b_0", "b_A", "b_P", "b_C",
          "sigma", "sigma_A", "sigma_P", "sigma_C")

Model_RE <- stan_model(file= "RE_Normal_APC.stan")

Result_RE <- sampling(
  Model_RE, data = StanData, pars = Pars, seed = 1234,
  chains = 4, iter = 6000, warmup = 1000, thin = 5,
  control = list(adapt_delta = 0.94, max_treedepth = 12))



# Ridge regression model

StanData <- list(Y=Y, X=X, N=N, I=I, J=J, K=K, L=L)

Pars <- c("b_0", "b_A", "b_P", "b_C", "sigma", "lambda")

Model_RR <- stan_model(file= "RR_Normal_APC.stan")

Result_RR <- sampling(
  Model_RR, data = StanData, pars = Pars, seed = 1234,
  chains = 4, iter = 6000, warmup = 1000, thin = 5,
  control = list(adapt_delta = 0.94, max_treedepth = 12))



# Random walk model

StanData <- list(Y=Y, X=X, N=N, I=I, J=J, K=K, L=L)

Pars <- c("b_0", "b_A", "b_P", "b_C",
          "sigma", "sigma_A", "sigma_P", "sigma_C")

Model_RW <- stan_model(file= "RW_Normal_APC.stan")

Result_RW <- sampling(
  Model_RW, data = StanData, pars = Pars, seed = 1234,
  chains = 4, iter = 6000, warmup = 1000, thin = 5,
  control = list(adapt_delta = 0.94, max_treedepth = 12))
\end{lstlisting}

The following R codes show the artificial parameters
and the results of the three models.
\begin{lstlisting}
> round(beta_A, digits = 2)
 [1]  0.5  0.3  0.3  0.1  0.1 -0.1 -0.1 -0.3 -0.3 -0.5

> round(beta_P, digits = 2)
 [1] 0 0 0 0 0 0 0 0 0 0

> round(beta_C, digits = 2)
 [1] -0.95 -0.75 -0.75 -0.55 -0.55 -0.35 -0.35 -0.15 -0.15  0.05
[11]  0.05  0.25  0.25  0.45  0.45  0.65  0.65  0.85  0.85



> print(Result_RE)
Inference for Stan model: RE_Normal_APC.
4 chains, each with iter=6000; warmup=1000; thin=5;
post-warmup draws per chain=1000, total post-warmup draws=4000.

          mean se_mean   sd   2.5%    25%    50%    75%  97.5% n_eff Rhat
b_0       0.01    0.00 0.27  -0.53  -0.16   0.00   0.18   0.55  3317    1
b_A[1]    0.94    0.00 0.24   0.46   0.79   0.94   1.09   1.40  3148    1
b_A[2]    0.63    0.00 0.24   0.15   0.48   0.63   0.78   1.09  3162    1
b_A[3]    0.53    0.00 0.24   0.05   0.39   0.53   0.68   0.99  3149    1
b_A[4]    0.24    0.00 0.24  -0.23   0.10   0.25   0.39   0.71  3143    1
b_A[5]    0.14    0.00 0.24  -0.34  -0.01   0.14   0.28   0.60  3152    1
b_A[6]   -0.15    0.00 0.24  -0.62  -0.29  -0.15   0.00   0.32  3152    1
b_A[7]   -0.26    0.00 0.24  -0.73  -0.41  -0.26  -0.11   0.20  3137    1
b_A[8]   -0.57    0.00 0.24  -1.05  -0.72  -0.57  -0.42  -0.10  3142    1
b_A[9]   -0.65    0.00 0.24  -1.13  -0.80  -0.65  -0.51  -0.19  3139    1
b_A[10]  -0.95    0.00 0.24  -1.43  -1.10  -0.95  -0.80  -0.49  3151    1
b_P[1]   -0.45    0.00 0.12  -0.68  -0.52  -0.45  -0.38  -0.21  3601    1
b_P[2]   -0.35    0.00 0.12  -0.59  -0.42  -0.35  -0.28  -0.11  3603    1
b_P[3]   -0.24    0.00 0.12  -0.48  -0.31  -0.24  -0.17   0.00  3594    1
b_P[4]   -0.15    0.00 0.12  -0.39  -0.22  -0.15  -0.08   0.08  3606    1
b_P[5]   -0.05    0.00 0.12  -0.29  -0.13  -0.06   0.01   0.18  3595    1
b_P[6]    0.04    0.00 0.12  -0.20  -0.03   0.04   0.11   0.27  3580    1
b_P[7]    0.15    0.00 0.12  -0.09   0.08   0.15   0.22   0.39  3571    1
b_P[8]    0.25    0.00 0.12   0.02   0.18   0.25   0.32   0.49  3577    1
b_P[9]    0.35    0.00 0.12   0.11   0.27   0.34   0.42   0.58  3589    1
b_P[10]   0.47    0.00 0.12   0.23   0.39   0.46   0.54   0.70  3581    1
b_C[1]   -0.06    0.00 0.04  -0.14  -0.09  -0.06  -0.04   0.01  3943    1
b_C[2]    0.06    0.00 0.03  -0.01   0.03   0.06   0.08   0.12  3798    1
b_C[3]   -0.05    0.00 0.03  -0.11  -0.07  -0.05  -0.03   0.00  3751    1
b_C[4]    0.04    0.00 0.03  -0.01   0.02   0.04   0.06   0.09  3907    1
b_C[5]   -0.05    0.00 0.02  -0.10  -0.07  -0.05  -0.04  -0.01  3859    1
b_C[6]    0.06    0.00 0.02   0.02   0.05   0.06   0.08   0.11  3429    1
b_C[7]   -0.05    0.00 0.02  -0.09  -0.06  -0.05  -0.04  -0.01  3683    1
b_C[8]    0.03    0.00 0.02  -0.01   0.02   0.03   0.04   0.07  3771    1
b_C[9]   -0.05    0.00 0.02  -0.08  -0.06  -0.05  -0.04  -0.01  3779    1
b_C[10]   0.05    0.00 0.02   0.01   0.04   0.05   0.06   0.08  3899    1
b_C[11]  -0.05    0.00 0.02  -0.09  -0.06  -0.05  -0.04  -0.02  3425    1
b_C[12]   0.06    0.00 0.02   0.02   0.04   0.06   0.07   0.09  3624    1
b_C[13]  -0.02    0.00 0.02  -0.06  -0.04  -0.02  -0.01   0.02  3268    1
b_C[14]   0.07    0.00 0.02   0.03   0.06   0.07   0.08   0.11  3859    1
b_C[15]  -0.04    0.00 0.02  -0.08  -0.05  -0.04  -0.02   0.01  3933    1
b_C[16]   0.06    0.00 0.03   0.00   0.04   0.06   0.07   0.11  3705    1
b_C[17]  -0.02    0.00 0.03  -0.08  -0.04  -0.02   0.00   0.03  3499    1
b_C[18]   0.00    0.00 0.03  -0.06  -0.02   0.00   0.03   0.07  3876    1
b_C[19]  -0.02    0.00 0.04  -0.09  -0.05  -0.02   0.00   0.05  3795    1
sigma     0.10    0.00 0.00   0.10   0.10   0.10   0.10   0.10  4056    1
sigma_A   0.73    0.00 0.22   0.44   0.58   0.68   0.82   1.27  3050    1
sigma_P   0.35    0.00 0.10   0.22   0.28   0.33   0.40   0.61  3170    1
sigma_C   0.06    0.00 0.01   0.05   0.05   0.06   0.07   0.09  3413    1
lp__    823.09    0.11 6.17 810.10 819.13 823.45 827.35 834.34  2903    1



> print(Result_RR)
Inference for Stan model: RR_Normal_APC.
4 chains, each with iter=6000; warmup=1000; thin=5;
post-warmup draws per chain=1000, total post-warmup draws=4000.

          mean se_mean   sd   2.5%    25%    50%    75%  97.5% n_eff Rhat
b_0       0.00    0.00 0.17  -0.33  -0.12  -0.01   0.11   0.32  2866    1
b_A[1]    0.80    0.00 0.12   0.56   0.72   0.80   0.88   1.03  3315    1
b_A[2]    0.52    0.00 0.12   0.29   0.45   0.53   0.60   0.75  3259    1
b_A[3]    0.46    0.00 0.11   0.24   0.38   0.46   0.53   0.67  3175    1
b_A[4]    0.20    0.00 0.11  -0.01   0.13   0.21   0.28   0.42  3180    1
b_A[5]    0.13    0.00 0.11  -0.08   0.06   0.13   0.20   0.34  3170    1
b_A[6]   -0.12    0.00 0.11  -0.33  -0.19  -0.12  -0.05   0.09  3237    1
b_A[7]   -0.20    0.00 0.11  -0.42  -0.28  -0.20  -0.13   0.01  3299    1
b_A[8]   -0.48    0.00 0.11  -0.70  -0.55  -0.48  -0.40  -0.25  3306    1
b_A[9]   -0.53    0.00 0.12  -0.75  -0.60  -0.53  -0.45  -0.29  3401    1
b_A[10]  -0.79    0.00 0.12  -1.02  -0.87  -0.79  -0.71  -0.54  3464    1
b_P[1]   -0.30    0.00 0.12  -0.54  -0.37  -0.29  -0.21  -0.07  3142    1
b_P[2]   -0.23    0.00 0.11  -0.46  -0.31  -0.23  -0.16  -0.02  3076    1
b_P[3]   -0.15    0.00 0.11  -0.37  -0.23  -0.15  -0.08   0.05  2914    1
b_P[4]   -0.10    0.00 0.10  -0.32  -0.17  -0.10  -0.03   0.10  2828    1
b_P[5]   -0.04    0.00 0.10  -0.25  -0.11  -0.04   0.03   0.16  2687    1
b_P[6]    0.02    0.00 0.10  -0.19  -0.05   0.02   0.09   0.22  2717    1
b_P[7]    0.10    0.00 0.10  -0.11   0.03   0.10   0.17   0.30  2602    1
b_P[8]    0.16    0.00 0.11  -0.05   0.09   0.16   0.24   0.38  2771    1
b_P[9]    0.23    0.00 0.11   0.01   0.16   0.23   0.30   0.45  2704    1
b_P[10]   0.32    0.00 0.12   0.09   0.24   0.32   0.39   0.54  2820    1
b_C[1]   -0.38    0.00 0.14  -0.64  -0.47  -0.38  -0.29  -0.11  3335    1
b_C[2]   -0.20    0.00 0.13  -0.45  -0.29  -0.20  -0.12   0.04  3297    1
b_C[3]   -0.29    0.00 0.12  -0.52  -0.37  -0.29  -0.21  -0.06  3221    1
b_C[4]   -0.16    0.00 0.11  -0.37  -0.23  -0.16  -0.08   0.05  3172    1
b_C[5]   -0.22    0.00 0.10  -0.42  -0.29  -0.22  -0.15  -0.03  3105    1
b_C[6]   -0.07    0.00 0.09  -0.25  -0.13  -0.07   0.00   0.11  3013    1
b_C[7]   -0.15    0.00 0.09  -0.32  -0.21  -0.15  -0.09   0.02  2912    1
b_C[8]   -0.03    0.00 0.08  -0.19  -0.09  -0.03   0.02   0.13  2876    1
b_C[9]   -0.08    0.00 0.08  -0.23  -0.13  -0.08  -0.03   0.07  2863    1
b_C[10]   0.05    0.00 0.08  -0.09   0.00   0.05   0.10   0.20  2892    1
b_C[11]  -0.02    0.00 0.08  -0.17  -0.07  -0.02   0.03   0.14  2908    1
b_C[12]   0.13    0.00 0.08  -0.03   0.07   0.12   0.18   0.28  3010    1
b_C[13]   0.08    0.00 0.09  -0.08   0.02   0.08   0.13   0.25  3047    1
b_C[14]   0.21    0.00 0.09   0.03   0.15   0.21   0.27   0.39  3103    1
b_C[15]   0.13    0.00 0.10  -0.06   0.06   0.13   0.20   0.33  3176    1
b_C[16]   0.26    0.00 0.11   0.06   0.19   0.26   0.33   0.47  3215    1
b_C[17]   0.21    0.00 0.12  -0.01   0.13   0.21   0.28   0.44  3263    1
b_C[18]   0.27    0.00 0.13   0.03   0.19   0.27   0.35   0.52  3336    1
b_C[19]   0.27    0.00 0.14   0.01   0.18   0.27   0.36   0.54  3367    1
sigma     0.10    0.00 0.00   0.10   0.10   0.10   0.10   0.10  3985    1
lambda    0.33    0.00 0.04   0.26   0.30   0.32   0.35   0.42  2680    1
lp__    825.80    0.12 6.17 812.68 821.83 826.07 830.14 836.98  2689    1



> print(Result_RW)
Inference for Stan model: RW_Normal_APC.
4 chains, each with iter=6000; warmup=1000; thin=5;
post-warmup draws per chain=1000, total post-warmup draws=4000.

          mean se_mean   sd   2.5%    25%    50%    75%  97.5% n_eff Rhat
b_0       0.00    0.00 0.00  -0.01  -0.01   0.00   0.00   0.01  3804    1
b_A[1]    0.50    0.00 0.02   0.46   0.49   0.50   0.51   0.53  3776    1
b_A[2]    0.29    0.00 0.01   0.26   0.28   0.29   0.30   0.32  3986    1
b_A[3]    0.29    0.00 0.01   0.27   0.28   0.29   0.30   0.32  3692    1
b_A[4]    0.11    0.00 0.01   0.09   0.10   0.11   0.11   0.13  3897    1
b_A[5]    0.10    0.00 0.01   0.08   0.09   0.10   0.10   0.11  3921    1
b_A[6]   -0.08    0.00 0.01  -0.10  -0.09  -0.08  -0.08  -0.07  3952    1
b_A[7]   -0.10    0.00 0.01  -0.12  -0.11  -0.10  -0.10  -0.08  4254    1
b_A[8]   -0.31    0.00 0.01  -0.34  -0.32  -0.31  -0.30  -0.29  3847    1
b_A[9]   -0.29    0.00 0.02  -0.33  -0.30  -0.29  -0.29  -0.27  3988    1
b_A[10]  -0.49    0.00 0.02  -0.53  -0.50  -0.49  -0.48  -0.46  3923    1
b_P[1]    0.00    0.00 0.01  -0.03   0.00   0.00   0.01   0.03  3857    1
b_P[2]    0.00    0.00 0.01  -0.02   0.00   0.00   0.01   0.02  3716    1
b_P[3]    0.00    0.00 0.01  -0.02   0.00   0.00   0.01   0.02  3670    1
b_P[4]    0.00    0.00 0.01  -0.02   0.00   0.00   0.00   0.01  4065    1
b_P[5]    0.00    0.00 0.01  -0.02  -0.01   0.00   0.00   0.01  4105    1
b_P[6]    0.00    0.00 0.01  -0.02  -0.01   0.00   0.00   0.00  3671    1
b_P[7]    0.00    0.00 0.01  -0.02   0.00   0.00   0.00   0.01  3818    1
b_P[8]    0.00    0.00 0.01  -0.02   0.00   0.00   0.00   0.02  3772    1
b_P[9]    0.00    0.00 0.01  -0.02   0.00   0.00   0.00   0.03  3873    1
b_P[10]   0.01    0.00 0.01  -0.02   0.00   0.00   0.01   0.04  3748    1
b_C[1]   -0.97    0.00 0.04  -1.05  -0.99  -0.97  -0.94  -0.89  3647    1
b_C[2]   -0.74    0.00 0.03  -0.81  -0.76  -0.74  -0.72  -0.67  4042    1
b_C[3]   -0.75    0.00 0.03  -0.80  -0.77  -0.75  -0.73  -0.69  3907    1
b_C[4]   -0.56    0.00 0.02  -0.60  -0.57  -0.56  -0.54  -0.51  4024    1
b_C[5]   -0.55    0.00 0.02  -0.59  -0.57  -0.55  -0.54  -0.51  3690    1
b_C[6]   -0.34    0.00 0.02  -0.37  -0.35  -0.34  -0.33  -0.30  4071    1
b_C[7]   -0.35    0.00 0.02  -0.38  -0.36  -0.35  -0.34  -0.32  3907    1
b_C[8]   -0.17    0.00 0.01  -0.19  -0.18  -0.17  -0.16  -0.14  3962    1
b_C[9]   -0.15    0.00 0.01  -0.17  -0.16  -0.15  -0.14  -0.13  4332    1
b_C[10]   0.05    0.00 0.01   0.03   0.04   0.05   0.06   0.07  3574    1
b_C[11]   0.05    0.00 0.01   0.02   0.04   0.05   0.05   0.07  3648    1
b_C[12]   0.26    0.00 0.01   0.23   0.25   0.26   0.27   0.28  4107    1
b_C[13]   0.28    0.00 0.02   0.25   0.27   0.28   0.29   0.31  3809    1
b_C[14]   0.47    0.00 0.02   0.43   0.46   0.47   0.48   0.51  3980    1
b_C[15]   0.46    0.00 0.02   0.42   0.45   0.46   0.47   0.50  3839    1
b_C[16]   0.66    0.00 0.02   0.61   0.64   0.66   0.67   0.70  3961    1
b_C[17]   0.68    0.00 0.03   0.62   0.66   0.68   0.70   0.73  3953    1
b_C[18]   0.80    0.00 0.03   0.74   0.79   0.81   0.82   0.87  3955    1
b_C[19]   0.88    0.00 0.04   0.79   0.86   0.88   0.91   0.96  3990    1
sigma     0.10    0.00 0.00   0.10   0.10   0.10   0.10   0.10  4114    1
sigma_A   0.17    0.00 0.05   0.11   0.14   0.16   0.20   0.30  2380    1
sigma_P   0.01    0.00 0.01   0.00   0.00   0.01   0.01   0.02  3889    1
sigma_C   0.15    0.00 0.03   0.11   0.13   0.15   0.17   0.22  2188    1
lp__    824.69    0.12 6.10 811.83 820.91 824.99 828.87 835.51  2695    1
\end{lstlisting}

\end{document}